\documentclass[fleqn,usenatbib]{mnras}
\usepackage{newtxtext,newtxmath}
\usepackage[T1]{fontenc}
\DeclareRobustCommand{\VAN}[3]{#2}
\let\VANthebibliography\thebibliography
\def\thebibliography{\DeclareRobustCommand{\VAN}[3]{##3}\VANthebibliography}
\usepackage{graphicx}
\usepackage{amsmath}
\usepackage{comment}
\usepackage{tabularx}
\usepackage{CJK}
\let\oldAA\AA
\renewcommand{\AA}{\text{\normalfont\oldAA}}
\usepackage{ulem}
\usepackage{xcolor}
\newcommand{{\dd}}{{\mathrm{d}}}

\title[Cosmic acceleration forecasted by Ly$\alpha$ forest]{Forecasting cosmic acceleration measurements using the Lyman-$\alpha$ forest}
\author[C. Dong et al.]
{Chenxing Dong (董辰兴),$^{1}$\thanks{E-mail: dcx@ufl.edu}
Anthony Gonzalez,$^{1}$\thanks{E-mail: anthonyhg@ufl.edu}
Stephen Eikenberry,$^{1}$
Sarik Jeram,$^{1}$
\newauthor
Manunya Likamonsavad,$^{1}$
Jochen Liske,$^{2}$
Deno Stelter,$^{3}$
Amanda Townsend$^{4}$
\\
$^{1}$Department of Astronomy, University of Florida, 211 Bryant Space Science Center, Gainesville, FL 32611, USA\\
$^{2}$Hamburger Sternwarte, Universit{\"a}t Hamburg, Gojenbergsweg 112, 21029 Hamburg, Germany\\
$^{3}$Department of Astronomy \& Astrophysics, UC Santa Cruz, 1156 High Street, Santa Cruz, CA 95064, USA\\
$^{4}$Apache Point Observatory and New Mexico State University, P.O. Box 59, Sunspot, NM, 88349-0059, USA
}
\pubyear{2022}

\begin{document}
\begin{CJK*}{UTF8}{gbsn}
\label{firstpage}
\pagerange{\pageref{firstpage}--\pageref{lastpage}}
\maketitle

\begin{abstract}
We present results from end-to-end simulations of observations designed to constrain the rate of change in the expansion history of the Universe using the redshift drift of the Lyman-$\alpha$ forest absorption lines along the lines-of-sight toward bright quasars. For our simulations we take Lyman-$\alpha$ forest lines extracted from Keck/HIRES spectra of bright quasars at $z>3$, and compare the results from these real quasar spectra with mock spectra generated via Monte Carlo realizations. We use the results of these simulations to assess the potential for a dedicated observatory to detect redshift drift, and quantify the telescope and spectrograph requirements for these observations. Relative to Liske et al. (2008), two main refinements in the current work are inclusion of quasars from more recent catalogs and consideration of a realistic observing strategy for a dedicated redshift drift experiment that maximizes $\dot{v}/\sigma_{\dot{v}}$. We find that using a dedicated facility and our designed observing plan, the redshift drift can be detected at $3\sigma$ significance in 15 years with a 25~m telescope, given a spectrograph with long term stability with $R=50,000$ and 25\% total system efficiency. To achieve this significance, the optimal number of targets is four quasars, with observing time weighted based upon $\dot{v}/\sigma_{\dot{v}}$ and object visibility. This optimized strategy leads to a 9\% decrease in the telescope diameter or a 6\% decrease in the required time to achieve the same S/N as for the idealized case of uniformly distributing time to the same quasars.
\end{abstract}

\section{Introduction} \label{sec:intro}
The expansion rate of the Universe changes with time for any model with non-zero energy density. Gravity decelerates the expansion, while dark energy \cite[e.g.,][]{Riess98,Schmidt98,Perlmutter99} leads to acceleration at late times. The detailed expansion history of the Universe is therefore a fundamental cosmological observable. Kinematic observations of the acceleration provide a direct, model-independent record of the impact of the matter and dark energy over cosmic time. 
As the recessional velocities of galaxies are characterized by the redshift, the changes in those velocities over time are therefore measured by the redshift drift ($\dd z/\dd t$).

The idea of redshift drift was first introduced by \cite{Sandage62} and \cite{McVittie62}. 
The redshift drift of a galaxy at redshift $z$ is given by 
\begin{equation}\label{eq:dzdt}
\frac{\dd z}{\dd t_\mathrm{obs}} = (1+z)H_0 - H(z),
\end{equation}
where $t_\mathrm{obs}$ is the time of the observation and $H(z)$ is the Hubble parameter. 
For a standard $\Lambda$CDM model \cite[e.g.][]{Planck16}, the observed redshift drift is negative at $z\gtrsim2$ and positive at lower redshifts.
Direct measurement of the redshift drift can distinguish among different dark energy models. 
None of the existing dark energy probes (e.g. weak lensing, BAO, SN~Ia) probe only the kinematics, but rather also depend on geometry or properties of the density perturbations \citep{Liske08}.
Measuring the redshift drift provides a model-independent determination of the expansion history of the Universe.

Several classes of targets have been proposed for redshift drift experiments. One is HI 21 cm absorption lines from radio-bright galaxies at a range of redshifts \citep{Darling12}. The HI 21 cm approach benefits from narrow intrinsic line widths and the ability in principle to detect lines at any redshift. \cite{Kloeckner15} have presented forecasts for cosmological constraints from the Square Kilometre Array (SKA) with this approach, finding the frequency shift can be estimated to a precision of one percent with a $\sim12$ years baseline, using SKA Phase 2 with 20 deg$^2$ field of view and 30,000 deg$^2$ survey coverage.
However, SKA Phase 2 will not be operational until late 2020s and thus a redshift drift detection will not be expected until at least the early 2040s.
As an alternative at low-redshift, \cite{Kim15} propose to use oxygen doublet lines in star-forming galaxies. This approach benefits from differential measurements of the wavelengths of the doublet lines to minimize systematics, but with the corollary challenge of a significantly reduced wavelength baseline.  The third approach, upon which we focus in the current paper, relies on Lyman-$\alpha$ forest lines in the foregrounds of bright quasars. First proposed by \cite{Loeb98}, this approach has the advantage of simultaneously probing a range of redshifts. It is however only possible at $z\gtrsim2$ for ground-based optical telescopes, and thus directly probes the epoch of deceleration. 

Though the equation for redshift drift is concise, the reality of conducting actual redshift drift experiments is challenging. 
For a \cite{Planck16} cosmology, the maximum expected annual recessional velocity change is $\sim +0.26~\mathrm{cm~s}^{-1}$ during the late-time epoch of acceleration; while during the epoch of deceleration for a galaxy at $z=3~(5)$ the expected annual change is $\sim -0.27~(-0.77)~\mathrm{cm~s}^{-1}$. For comparison, the best exoplanet radial velocity experiments in the world (e.g. HARPS \citep{Pepe02, Mayor03} and NEID \citep{Schwab16, Logsdon18}) achieve 
precisions of $\sim10$ -- $30~\mathrm{cm~s}^{-1}$ with stability of weeks. Using laser frequency comb \citep{Wilken12} or Fabry-Perot etalon \citep{Schwab15} calibrations one can even achieve a precision of $\sim3~\mathrm{cm~s}^{-1}$ with stability of hours to days.
The best existing observational limit on redshift drift \citep{Darling12} is approximately two orders of magnitude larger than the expected signal.

Detection of redshift drift thus requires improvements in both precision and long-term stability of radial velocity measurements. For all of the proposed classes of targets, the instrument must be stable on decadal timescales. Moreover, the targets are generally faint and thus require large collecting areas. Proposed experiments in the literature include facilities such as the Extremely Large Telescope (ELT; \citealt{Liske08}), the Square Kilometer Array (SKA; \citealt{Kloeckner15}) and the Five-hundred-meter Aperture Spherical Telescope (FAST; \citealt{Jiao20}).

Besides instrumental requirements, requirements on the properties of the target population are also strict. First, the targets should trace the Hubble flow, and their non-cosmological velocities should be stable over the period of the experiment. Second, the targets should ideally have spectra with narrow lines, enabling the measurement of small velocity shifts with high significance.
Finally, within a given class of targets, the sample must be carefully selected to yield the optimal signal-to-noise ratio (S/N).

In this paper we revisit the analysis of \cite{Liske08} to explore the potential for a redshift drift experiment based upon the Lyman-$\alpha$ forest. In the subsequent decade after publication of that paper, a number of new bright QSOs have been discovered \citep[e.g.][]{Schindler17,Schindler19b,Boutsia20,Jeram20}, potentially improving the prospects for such an experiment. There has also been renewed interest in such an experiment driven by technological advances that may enable construction of a dedicated facility using arrays of small telescopes \citep{Eikenberry19a,Eikenberry19b,Eikenberry19c}.
We present end-to-end simulations to quantify the constraints that can be derived via the Ly$\alpha$ forest approach for a dedicated facility, updating the work of \cite{Liske08} in several ways. First, we use the most recent catalogs to provide forecasts for both northern and southern hemisphere facilities. 
Second, we revisit the comparison of \cite{Liske08} between simulated and real line lists using high-resolution spectra for a sample that includes QSOs that are best suited for such an experiment.
In this work we directly compare the results from this approach with those obtained using high-resolution spectra of real quasars. Third, our simulations include practical observing strategies based upon the visibilities of individual quasars for a dedicated facility. With these modifications, we estimate the time necessary to detect redshift drift.

In Section \ref{sec:genspec} we describe our methods in generating spectral templates that can be used in simulating Ly$\alpha$ forest observations at different time periods.
In Section \ref{sec:uncertainty} we present the simulated process of obtaining observed spectra at different epochs, calculating redshift drifts and the uncertainties associated with them.
In Section \ref{sec:experiment} we discuss aspects of a redshift drift experiment in terms of target selection, spectral template generation method selection, uncertainty dependences on redshift and spectral resolution, observing time allocation and prospects for a redshift drift detection.
Our conclusions are summarized in Section \ref{sec:conclusion}.

\section{Spectral template generation}\label{sec:genspec}
The first key step in our simulations is generating spectral templates, which are parameterized Ly$\alpha$ forest spectra without any noise. We take two approaches to template generation, so that a cross-comparison allows us to quantify the consistency of the two methods. First, we use archival Keck/HIRES spectra and fit the Ly$\alpha$ absorption lines to generate spectral templates directly from known bright quasars. Second, we use published analyses of the distributions of Ly$\alpha$ line properties as a function of redshift \citep{Kim01,Liske08} to generate synthetic spectral templates, as was done in \cite{Liske08}. 
The latter approach has the advantages of enabling simulations of QSOs at arbitrary redshifts with arbitrary magnitudes, and also more computationally efficient in line profile fitting on the spectra.
The former has the advantage of more directly mimicking the actual observations, particularly in instances where the HIRES archive includes spectra for QSOs that would be included in the target list for a redshift drift experiment.  

\subsection{Keck HIRES Spectroscopy}
The Keck Observatory Archive \citep[KOA;][]{Berriman05,Berriman14} contains spectra from the High Resolution Echelle Spectrometer \citep[HIRES;][]{Vogt94} with $R>25,000$ for a number of bright quasars.
For each target in our sample (defined in Section \ref{sec:target}), we check if there are available spectra in the KOA.
For targets with multiple observations in the KOA, we choose the observation with the best data. 
We then calibrate the data and normalize the continua as described in Section \ref{sec:keck}, and fit line profiles in the Ly$\alpha$ forest as described in Section \ref{sec:voigtfit} to form the lists of line parameters.

\subsubsection{Data Preprocessing}\label{sec:keck}
The HIRES data require several steps of pre-processing before line fitting. First, because HIRES is a high-resolution echelle spectrograph, the wavelength range of interest typically spans multiple spectral orders.  Within a given observation, for data with overlapping wavelengths, we choose those with the highest S/N in flux. We connect the selected pieces of data to form the spectra, and use the section between Ly$\beta$ and Ly$\alpha$ as the Ly$\alpha$ forest for our analysis.

\begin{figure*}
\includegraphics[width=\textwidth]{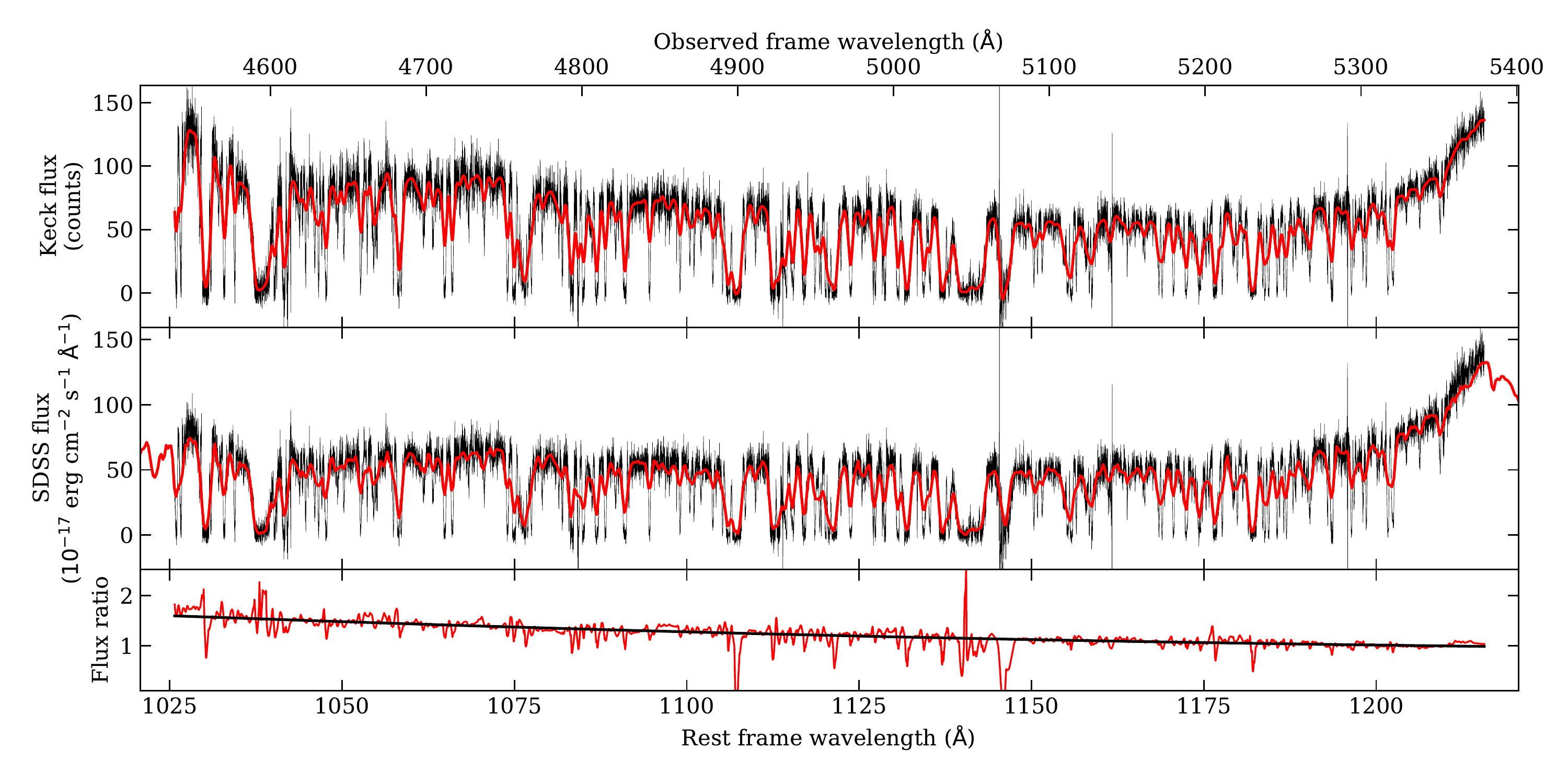}
\caption{
For the QSO SDSS~J173352.23+540030.4, shown here are both the Keck HIRES spectra and the SDSS data used for flux calibration. In the top panel the black spectrum is the raw Keck/HIRES data, while the red curve is the same data smoothed to SDSS resolution. In the middle panel we show the SDSS spectrum in red and the Keck spectrum after it has been flux calibrated using the SDSS spectrum in black. The bottom panel illustrates the applied correction used to transform the HIRES spectrum. The red curve is the ratio between the smoothed HIRES spectrum and the SDSS spectrum. We also show the 4th order polynomial fit to the median-filtered ratio (black).
}\label{fig:fluxcal}
\end{figure*}

Because the HIRES spectra are not flux-calibrated, we must apply a flux-calibration from another observational source. 
The flux calibration is needed for modeling the photon noise of the generated spectra.
For quasars with available SDSS \citep[e.g.][]{Blanton17} spectra, we smooth the HIRES spectra to the SDSS resolution, compute the flux ratio between the smoothed HIRES spectra and the SDSS spectra, smooth the ratio with a median filter, fit polynomials to the ratio and scale the original HIRES spectra by the polynomial-fitted flux ratio. A demonstration of the flux calibration using SDSS spectra is shown in Figure~\ref{fig:fluxcal}.
We note that there are up to 5\% calibration errors in the SDSS flux \citep[e.g.][]{Bautista17}, but the calibrated flux is only used in modeling the photon noise of the generated spectra, and this level of calibration error has a negligible impact on our analysis. The Voigt profile fitting and spectral template generation are not affected by the flux calibration.
Because there is no SDSS data for B1422+231 and S5~0014+81 (see Section \ref{sec:target}), we instead scale the spectrum to match the flux at rest frame $1450~\mathrm{\AA}$ to the quasars'
absolute magnitude at rest-frame $1450~\mathrm{\AA}$ (hereafter $M_{1450}$), which is a commonly used metric for a quasar's luminosity due to lack of line features at this wavelength.
Given that $M_{1450}$ is only used for two quasars, the additional uncertainty does not qualitatively impact our results.

\begin{figure*}
\includegraphics[width=\textwidth]{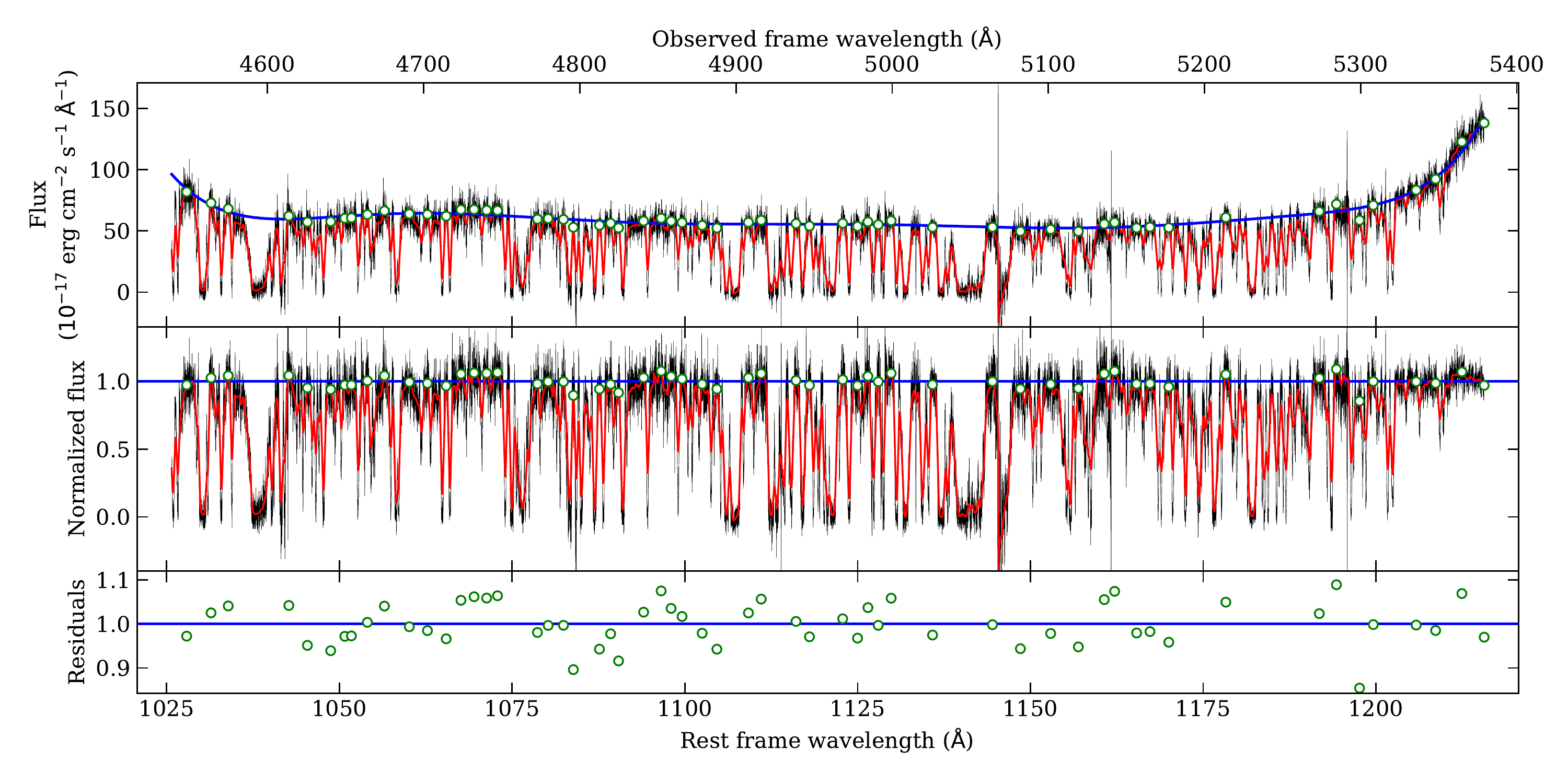}
\caption{Illustration of the normalization of the spectrum prior to line fitting. The top panel shows the flux-calibrated spectrum (black), and the same spectrum after smoothing with a Gaussian kernel ($\sigma=15$ pix, red). The green circles correspond to local maxima in the smoothed spectrum, with a minimum distance of 100 pixels between maxima, which are used to derive a continuum model. The blue curve is the best-fit polynomial representation of the continuum. In the middle panel we show the same spectrum, but now divided by the polynomial continuum model such that the continuum is normalized to 1. It is from this spectrum that we identify absorption lines. The green circles again denote the local maxima used to define the continuum. The lower panel shows the fractional residuals of the local maxima.
}\label{fig:contsub}
\end{figure*}

We next perform a continuum normalization for each spectrum. We smooth the raw spectra with a Gaussian kernel ($\sigma=15$ pixel). 
The kernel size was a selected as a trade-off, as oversmoothing will lead to an underestimation of the continuum level, while too small a kernel will fail to smooth out local fluctuations. Our results are not strongly sensitive to the precise value; kernels of 10 or 20 pixels yield similar results.
To distinguish the continuum from absorption lines, we identify local flux maxima in the smoothed spectra, limiting the minimum distance between input maxima to be $\ge 100$ pixels.
This minimum distance is somewhat arbitrary, but our results are again relatively insensitive to the precise value as long as it is sufficiently small to preserve the large-scale structure of the continuum and sufficiently large to avoid underfitting the continuum due to the inclusion of local maxima.
We visually inspect the selected local maxima, remove those that are inconsistent with tracing the continuum level, and add more points as needed. 
We then fit polynomials of order 10 for a spectral width of $\sim 1000$ Angstroms, and divide by the best fit continua to yield output spectra that are renormalized to unity.
The result for a typical continuum fit is shown in Figure~\ref{fig:contsub}. As can be seen by the residuals, the polynomial removes the general continuum structure without impacting the depth of the individual lines.

\subsubsection{Voigt profile fitting}\label{sec:voigtfit}
We use the continuum-subtracted HIRES spectra as input to identify and fit Ly$\alpha$ forest lines. We first identify candidate lines using
the crude significance level (CSL; \citealt{Gaikwad17}) and set the line detection threshold to be $\mathrm{CSL} \ge1.5$. We divide the spectra into small regions based on line locations and fit Voigt profiles in each region. 
Details of the equations used for the Voigt profile fitting are given in Section \ref{sec:voigteqs} in the appendix.

We use the {\scriptsize\texttt{LMFIT}} package \citep{lmfit} to do least-square fitting.
During fitting, we set the minimum line width to be $b_{\mathrm{min}} = 5~\mathrm{km~s}^{-1}$ to remove artifacts.
After an initial fit of each region, we decided whether to add another line using the Akaike Information Criterion with Correction \citep[AICC;][]{Akaike74,Liddle07,Gaikwad17}. After two consecutive failed attempts trying to add lines, the algorithm moves on to the next region.
An example section of the results of the Voigt profiles fitting is shown in the top panel of Figure~\ref{fig:linelist}.

We used the fitted line parameters to generate the spectral templates for our experiment. For some HIRES spectra where there are small gaps in the wavelength space ($< 25~\mathrm{\AA}$ in rest frame), we filled in the gaps by sampling from existing lines with the average line densities of the corresponding spectra.

\subsection{Synthetic spectral template}
In addition to deriving line lists from real quasar spectral templates, we also pursue the alternative approach of generating synthetic spectra based upon published line parameter distributions. 
We use this approach to compare our results with \cite{Liske08}, and we also test the degree to which the two independent approaches are consistent for our analysis.

\begin{figure*}
    \includegraphics[width=\textwidth]{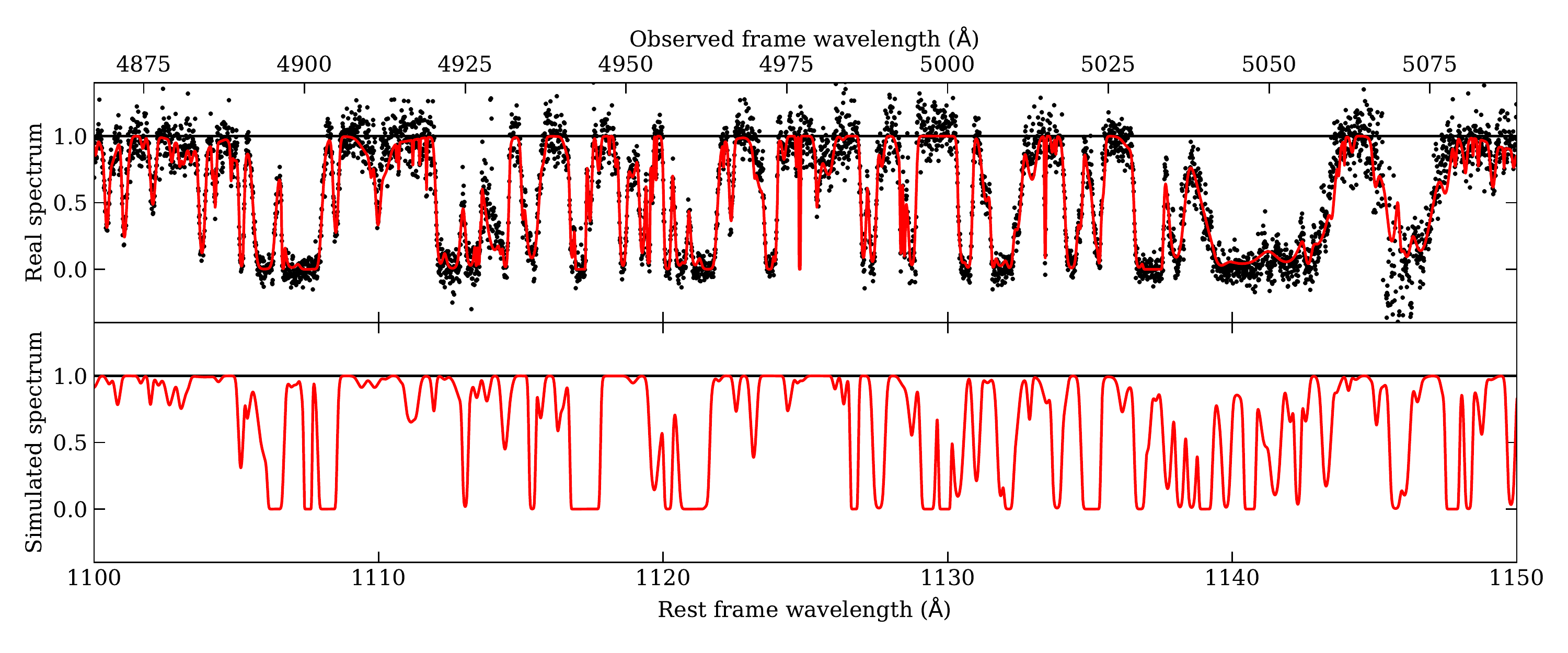}
    \caption{
Shown above is a subregion ($\Delta\lambda=200$~\r{A}) of the Ly$\alpha$ forest spectrum for a $z=3.4$ QSO. The top panel shows in red the spectrum generated by fitting Keck data (black points), while the lower panel is generated by randomly drawing line parameters from published empirical distributions. In these panels we have added no noise to facilitate visual comparison.
}\label{fig:linelist}
\end{figure*}

The generated line lists are represented by lists of Voigt profiles as described in Section \ref{sec:voigtfit}.
We obtain empirical parameter distributions for the Doppler parameter $b$, the HI column density $N_\mathrm{HI}$ and the redshifts $z$ of each absorption line to form the Ly$\alpha$ forest \citep{Kim01,Liske08}.
The distributions of these parameters are 
$\dd n/\dd b \sim \exp\left(-b_\sigma^4/b^4\right)/b^5$ where $b_\sigma=27.75$ $\mathrm{km~s}^{-1}$ for a mean redshift of $\langle z\rangle = 2.87$ \citep{Hui99}, 
$\dd n/\dd N_\mathrm{HI} \sim N_\mathrm{HI}^{-1.46}$ at $z\sim 2.8$ \citep{Hu95}, and $\dd n/\dd z = (9.06 \pm 0.40) (1+z)^{2.19 \pm 0.27}$ for $N_\mathrm{HI}$ in the range $[10^{13.64},10^{16}]$ $\mathrm{cm}^{-2}$ \citep{Kim01}. 
We integrate the fiducial $\dd n/\dd z$ distribution from Ly$\beta$ to Ly$\alpha$ to get the number of lines for $13.64<\log N_\mathrm{HI} (\mathrm{cm}^{-2})<16$, and convert it to the number of lines for $12<\log N_\mathrm{HI} (\mathrm{cm}^{-2})<16$ using the $\dd n/\dd N_\mathrm{HI}$ distribution. 
The Doppler parameter, column density and redshift are generated independently, as was the case in \cite{Liske08}. While treating these quantities as independent is a simplification, simulations have shown that there is large scatter and minimal correlation between the two parameters \citep[e.g.][]{Zhang98}. \cite{Kim01} also treated the distributions for these two parameters as independent except for a $b$ cutoff.
An example section of our generated spectral template is shown in the bottom panel of Figure~\ref{fig:linelist}.

\section{Redshift Uncertainty Estimation}\label{sec:uncertainty}
Using the generated spectral templates, we simulated observed spectra at different epochs, derive the estimated redshift drift from the observations, and estimate the associated uncertainties.

\subsection{Simulating Redshift Drift}
Simulating a redshift drift experiment requires that we must simulate the spectra at a number of different epochs using realistic observational parameters. 
We use a baseline resolution of $R = 50,000$ and a three-pixel resolution element, corresponding to a pixel size of $\delta \lambda = 0.0333$~\r{A} at 5000~\r{A}. 
We generate spectra at the appropriate resolution using either the Keck line lists or the line lists generated from the empirical line parameter distributions.

To form a second epoch spectrum with redshift drift $\delta z$, given by Equation \ref{eq:dzdt} for a fiducial input cosmology, we add $\delta \lambda = \frac{\lambda_0}{1+z}\delta z$ to the line center parameters $\lambda_0$ of the Ly$\alpha$ lines in our generated line lists, and change the line width parameter $b$ in the same way.
The pixel scale and resolution for the second epoch spectra are set the same as described above.

We assume the experiment is photon noise limited, because we calculated and found that for such observations the photon noise greatly exceeds readout noise and dark current, even for the smallest apertures under consideration, or for the absorption regions where flux is low.
We add Poisson noise to the spectra at each epoch, given an observational setup (telescope aperture, exposure time and total efficiency). 
Given the flux density $F_\lambda$ at each pixel, the observed wavelength $\lambda_\mathrm{obs}$, the spectral dispersion per pixel $\delta \lambda$, the telescope aperture $D$, integration time $t_\mathrm{int}$ and efficiency $\epsilon$, the number of photons collected for each pixel can be estimated by
\begin{multline}
N_\mathrm{phot} = 1.20\times 10^6
\frac{F_\lambda}{10^{-15}\ \mathrm{erg}\ \mathrm{cm}^{-2}\ \mathrm{s}^{-1}\ \mathrm{\AA}^{-1}}
\frac{\lambda_\mathrm{obs}}{5000\ \mathrm{\AA}}\\
\frac{\delta \lambda}{0.03\mathrm{\AA}}
\left(\frac{D}{15\ \mathrm{m}}\right)^2
\frac{t_\mathrm{int}}{100\ \mathrm{h}}
\frac{\epsilon}{0.25}.
\end{multline}

\subsection{Recovering velocity drift}
We extract velocity information from the observed spectra following \cite{Bouchy01} and \cite{Liske08}, where each pixel of the spectra is considered independent.
Given a set of spectra observed at multiple epochs, assume the time interval from the beginning of the experiment to the $j$th epoch is $\Delta t_j$, and the flux at the $i$th pixel of the spectrum observed at the $j$th epoch is $S_{ij}$, with an uncertainty of $\sigma_{ij}$.
Then the average velocity drift for the $i$th pixel can be estimated as
\begin{equation}
\dot{v}_i = \frac{m_ic}{\lambda_i S_{i}^{\prime}},
\end{equation}
where $S_{i}^{\prime}\equiv \dd S_i/\dd \lambda$ is the slope of the spectrum at the $i$th pixel, which is assumed to be constant over different epochs, and
\begin{equation}
m_i \equiv \frac{
    \overline{S_i\Delta t} - \overline{S_i}\ \overline{\Delta t}
}{
    \overline{\Delta t^{2}}-\overline{\Delta t}^{2}
},
\end{equation}
where $S_i$ and $\Delta t$ represent the set $\{S_{ij}\}$ and $\{\Delta t_j\}$ for all $N_e$ epochs, and the bar denotes the weighted average over all epochs:
\begin{equation}
\overline{x}\equiv \frac{\sum_{j=1}^{N_{\mathrm{e}}} x \sigma_{ij}^{-2}}{\sum_{j=1}^{N_{\mathrm{e}}} \sigma_{ij}^{-2}}.
\end{equation}
For real experiments it may be more optimal to use a weighting scheme that is robust to outliers, but inverse variance weighting is sufficient for the current analysis.
The uncertainty in $\dot{v}_i$ can be estimated as
\begin{equation}
\sigma_{\dot{v}_i}^{2}=\left(\frac{c}{\lambda_i S_{i}^{\prime}}\right)^{2}
\left(\sigma_{m_{i}}^{2}+m_{i}^{2}\frac{\sigma_{S_{i}^{\prime}}^{2}}{{S_{i}^{\prime}}^{2}} \right),
\end{equation}
where $\sigma_{S_{i}^{\prime}}$ is the uncertainty in the slope of the spectrum, which is propagated from the uncertainty in the flux, and
\begin{equation}
\sigma_{m_{i}}^{2}=\left[\sum_{j} \sigma_{ij}^{-2}\left(\overline{\Delta t^{2}}-\overline{\Delta t}^{2}\right)\right]^{-1}
\end{equation}
is the variance of $m_i$. Finally, the overall uncertainty $\sigma_{\dot{v}}$ for all pixels is the weighted average of $\sigma_{\dot{v}_i}$:
\begin{equation}\label{eq:overall_sigma}
\sigma_{\dot{v}}^{2}=\frac{\sum_{i} \sigma_{\dot{v}_i}^{2} w_{i}^{2}}{\left(\sum_{i} w_{i}\right)^{2}}=\frac{1}{\sum_{i} \sigma_{\dot{v}_i}^{-2}}.
\end{equation}

\section{Experiment Simulation}\label{sec:experiment}
In this section, we apply the previously discussed method to simulated observations of selected quasars and discuss the uncertainty in our experiment and the prospects for a detection.

\subsection{Target Selection}\label{sec:target}

\begin{table*}
\caption{Target selection for the northern sample.}
\label{tab:qsos}
\begin{tabular}{clrrccc}
\hline
Rank & Name & RA (deg)&     DEC (deg) & $M_\mathrm{1450}$ & $r$ (mag) & $z$\\
\hline
\multicolumn{6}{c}{Selected by $|\dot{v}|/\sigma_{\dot{v}}$}\\
\hline
1  &           {\bf APM 08279+5255} &  127.923758 &  52.754866 & -30.96 &  15.32 &  3.911 \\
2  &                {\bf B1422+231} &  216.158784 &  22.933531 & -30.31 &  15.31 &  3.620 \\
3  &  {\bf PS1 J212540.96-171951.4} &  321.420691 & -17.330947 & -29.35 &  16.50 &  3.900 \\
4  &           {\bf PSS J0926+3055} &  141.651381 &  30.918037 & -29.00 &  17.06 &  4.190 \\
5  & {\bf SDSS J034151.16+172049.7} &   55.463213 &  17.347164 & -29.59 &  16.26 &  3.690 \\
6  &                 PSS J1723+2243 &  260.846281 &  22.732538 & -28.49 &  18.06 &  4.520 \\
7  &       SDSS J161737.78+595020.1 &  244.407437 &  59.838926 & -28.62 &  17.74 &  4.315 \\
8  &        PS1 J052136.92-133938.8 &   80.403846 & -13.660775 & -28.36 &  17.68 &  4.270 \\
9  &       SDSS J164804.84+493326.7 &  252.020179 &  49.557447 & -28.73 &  17.62 &  4.220 \\
10 &       SDSS J095937.11+131215.5 &  149.904628 &  13.204312 & -28.51 &  17.39 &  4.064 \\
\hline
\multicolumn{6}{c}{Selected by $M_\mathrm{1450}$}\\
\hline
1  &           {\bf APM 08279+5255} &  127.923758 &  52.754866 & -30.96 &  15.32 &  3.911 \\
2  &                {\bf B1422+231} &  216.158784 &  22.933531 & -30.31 &  15.31 &  3.620 \\
3  &                     S5 0014+81 &    4.285417 &  81.585556 & -29.65 &  16.30 &  3.366 \\
4  & {\bf SDSS J034151.16+172049.7} &   55.463213 &  17.347164 & -29.59 &  16.26 &  3.690 \\
5  &  {\bf PS1 J212540.96-171951.4} &  321.420691 & -17.330947 & -29.35 &  16.50 &  3.900 \\
6  &                   HS 1946+7658 &  296.228556 &  77.097965 & -29.18 &  16.17 &  3.051 \\
7  &                 PMN J1451-1512 &  222.946046 & -15.205618 & -29.14 &  18.73 &  4.763 \\
8  &                   SBS 1425+606 &  216.734116 &  60.430785 & -29.13 &  16.24 &  3.186 \\
9  &                   HS 0741+4741 &  116.340783 &  47.576720 & -29.13 &  16.26 &  3.220 \\
10 &           {\bf PSS J0926+3055} &  141.651381 &  30.918037 & -29.00 &  17.06 &  4.190 \\
\hline
\multicolumn{6}{l}{The overlapping QSOs between the two selections are labeled by bold names.}
\end{tabular}
\end{table*}

The last decade has seen a significant increase in the number of bright QSOs known at high redshift.  
We aim here to identify the best targets for both northern and southern hemisphere facilities. 
We select our targets mostly from three published catalogs of luminous quasars. The first catalog is the Extremely Luminous Quasar Survey (ELQS) in the Sloan Digital Sky Survey (SDSS) Footprint catalog of 407 quasars \citep{Schindler17,Schindler18,Schindler19a}, with $i$ band magnitude $m_i < 18.5$ and $z > 2.5$, reaching a completeness of $>70\%$ in the core region of the survey ($3<z<5$; $m_i < 17.5$).
The second catalog is the ELQS in the Pan-STARRS 1 footprint catalog of 592 quasars (PS-ELQS; \citealt{Schindler19b}),
while 195 quasars are overlapped with ELQS. The PS-ELQS catalog is cut at $m_i < 18.5$ and $z > 2.8$, achieving a completeness of $>70\%$ for $3.5<z<5$ within the footprint of the survey covering $\sim21,486~\mathrm{deg}^2$.
The third catalog is the QUasars as BRIght beacons for Cosmology in the Southern Hemisphere (QUBRICS; \citealt{Calderone19}) catalog, with $m_i < 18$ and $z > 2.5$, reaching a completeness of $>90\%$.
From the QUBRICS catalog we collected the ``golden sample'' of 30 southern QSOs for the Sandage Test selected by \cite{Boutsia20}.
In addition to these three catalogs, we did a literature search to identify any QSOs at $z>2$ that are sufficiently luminous to warrant consideration, and found two --- S5~0014+81 \citep{Kuhr83} and HS~1946+7658 \citep{Hagen92}.

From this larger sample, we start by considering the 10 best targets in each hemisphere based upon the S/N of the expected redshift drift signal, $|\dot{v}|/\sigma_{\dot{v}}$, for a \cite{Planck16} cosmology in this paper. 
The choice of 10 targets as a starting point is designed to ensure a broad distribution in RA while acknowledging that the best S/N will always be achieved from focusing upon the best targets. In Section \ref{sec:prospect} we will address the question of the optimal sample size for such an experiment.
As was demonstrated in \cite{Liske08}, selecting targets based upon their expected $|\dot{v}|/\sigma_{\dot{v}}$ is expected to yield a higher S/N for detecting redshift drift than a simple magnitude-limited sample, and this is the primary target selection approach in this paper. 
In $|\dot{v}|/\sigma_{\dot{v}}$ the $\dot{v}$ is based solely upon the input cosmology \citep{Planck16} and $\sigma_{\dot{v}}$ is based solely on the continuum flux level of the quasar. While specific features in the Lyman forest may vary between quasars,
the assumption of cosmological isotropy implies that all site lines will be effectively equivalent.
The continuum level is important, as this will drive the overall noise level.

We estimate the $|\dot{v}|/\sigma_{\dot{v}}$ for the quasars in the our sample. Following \cite{Liske08}, the uncertainty in the $\dot{v}$ measurement is $\sigma_{\dot{v}} \propto 10^{0.2r}(1+z_\mathrm{QSO})^{-1.7}$ for a given observation set-up, where $r$ is the SDSS $r$-band magnitude of the quasar, obtained from the SDSS or Pan-STARRS \citep{Panstarrs} catalog. The theoretical $\dot{v}$ is then calculated under \cite{Planck16} cosmology. 
We selected the northern (southern) sample by restricting the declinations to be greater than $-20^\circ$ (less than $+20^\circ$).
The northern $|\dot{v}|/\sigma_{\dot{v}}$ selected quasars are listed in the top panel of Table~\ref{tab:qsos}. 
For the southern $|\dot{v}|/\sigma_{\dot{v}}$ selected sample, the golden sample of southern QSOs for the Sandage Test from \cite{Boutsia20} was taken into consideration and one of their quasars (SMSS J054803.19-484813.1) was added to our sample. Our southern sample is presented in Section \ref{sec:southern} in the appendix.

For illustration, we also provide a brief comparison with a sample selected based purely upon $M_{1450}$.
The quasars were ranked by their $M_{1450}$ and the top 10 were selected from the sample. For the two additional quasars, the $M_{1450}$ of S5~0014+81 was estimated using the spectra from \cite{Kuhr83}; for HS~1946+7658, the quantity was estimated using \cite{Hernan16} composite spectra \citep{Jeram20}.
The northern sample selected by $M_{1450}$ is listed in bottom panel of Table~\ref{tab:qsos}. Five quasars are overlapping between the two selections, and are labeled bold in the table.

\begin{figure}
    \includegraphics[width=\columnwidth]{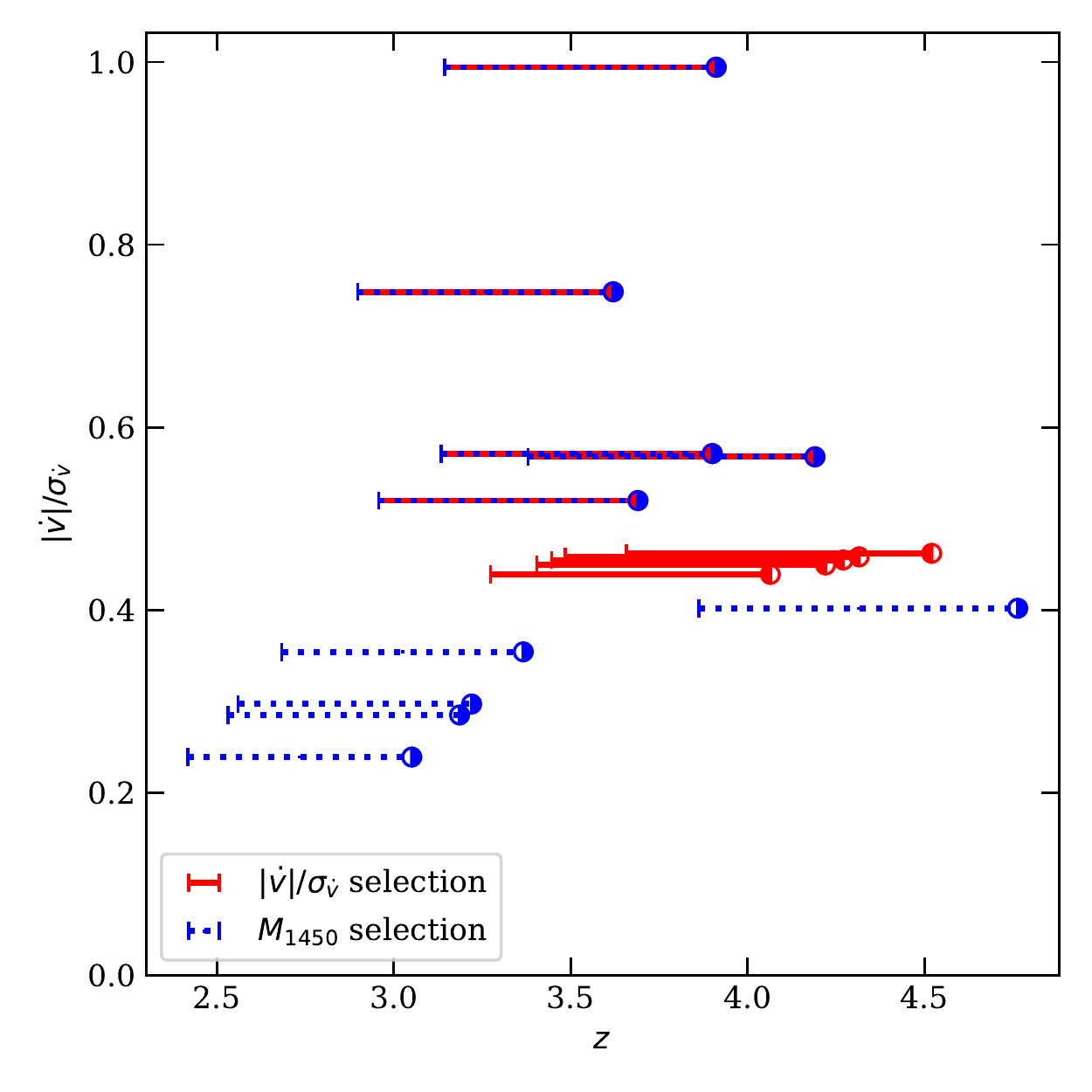}
    \caption{Comparison of the two target selection methods. Plotted is the predicted $|\dot{v}|/\sigma_{\dot{v}}$ for the $|\dot{v}|/\sigma_{\dot{v}}$  selection (red) and $M_{1450}$ selection (blue) as a function of redshift. The nodes indicate the redshift of the quasars and the bars indicate the redshift span of the corresponding Ly$\alpha$ forests.}\label{fig:targetselect}
\end{figure}

We compared the predicted $|\dot{v}|/\sigma_{\dot{v}}$ in both selections, 
assuming observing 50 hours per week for 20 years, and the observation time is equally distributed among the 10 quasars, and one spectrum is generated in every 2 months. 
A 15~m telescope and a total efficiency (telescope+instrument+sky) of 25\% is used. 
As shown in Figure~\ref{fig:targetselect}, the five targets overlapping between the two selection methods have the top five $|\dot{v}|/\sigma_{\dot{v}}$ predictions. The $|\dot{v}|/\sigma_{\dot{v}}$ selection has a lower limit of $|\dot{v}|/\sigma_{\dot{v}}$ at 0.44, while the $M_{1450}$ selection has a lower limit at 0.24. Therefore, we decided to use the $|\dot{v}|/\sigma_{\dot{v}}$ selection in the following analyses.

\subsection{Reconstructed versus synthetic spectral templates}\label{sec:realllvssimll}
\begin{figure*}
    \includegraphics[width=\columnwidth]{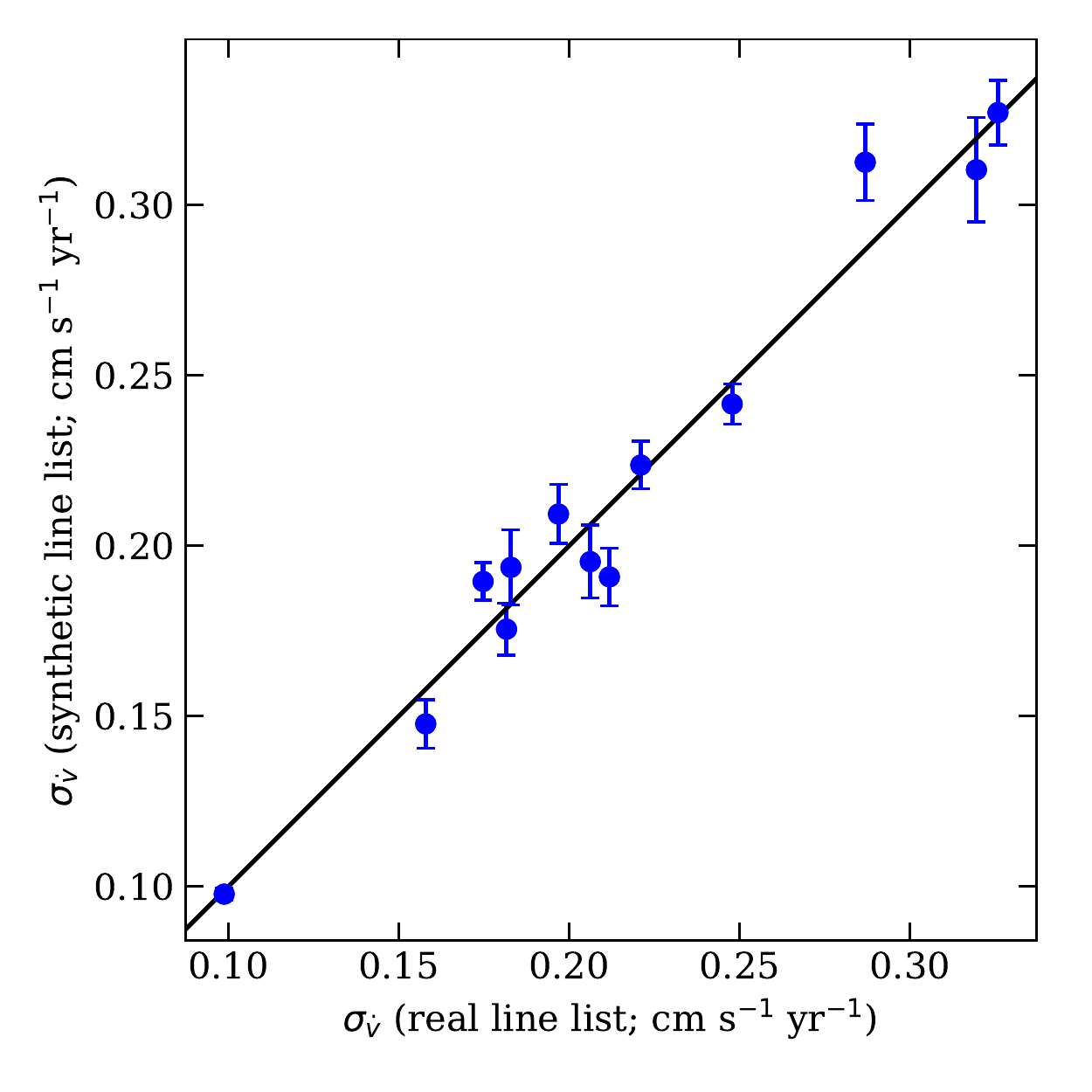}
    \includegraphics[width=\columnwidth]{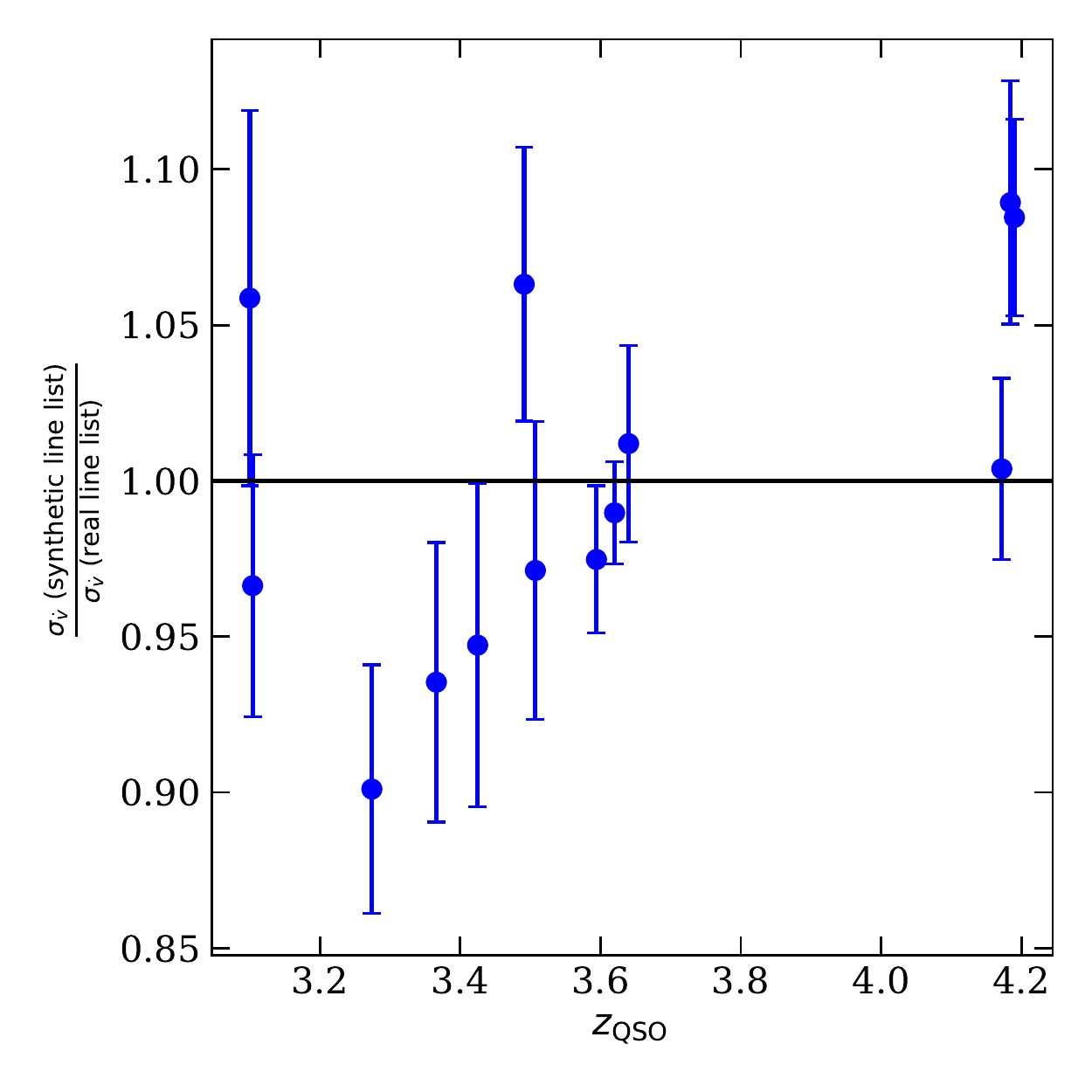}
    \caption{Comparison of the estimated uncertainties in velocity drift $\sigma_{\dot{v}}$ using reconstructed line lists versus synthetic line lists. The uncertainties are estimated using mock observations of each quasar for 50 hours per week for 20 years. Left: the one-to-one plot of the $\sigma_{\dot{v}}$ using the two line lists. The black line shows the 1:1 relation. Right: the ratio of $\sigma_{\dot{v}}$ between the two line list realizations as a function of the redshift of the corresponding quasars.
    }\label{fig:complinelist}
\end{figure*}

We generate Ly$\alpha$ forest line lists both from reconstructed spectra of real quasars and from literature line parameter distributions, as described in Section \ref{sec:genspec}. In this section, we check the consistency in the resulting redshift drift measurement between the two methods. 

We utilized 13 available reduced spectra to form the reconstructed line list, and also generated synthetic spectral templates given the redshifts of those 13 quasars. Both line lists were assigned the continuum flux of corresponding quasars. We generated mock observations of each quasar for 50 hours per week for 20 years. To simplify the computation, we assumed the observations took place only at the two end points of the 20-year period, and noted that distributing the observations into equally spaced nights would increase the estimated $\sigma_{\dot{v}}$ by a factor of $\sim 1.7$ \citep{Liske08}.
The mock observations utilized a 15~m telescope with an efficiency of 25\%.

The estimated uncertainties in velocity drift $\sigma_{\dot{v}}$ for these mock observations using both reconstructed line lists and synthetic line lists are shown in Figure~\ref{fig:complinelist}. The two line list realizations have comparable uncertainty estimations, with an rms scatter between the two of $0.012~\mathrm{cm~s}^{-1}~\mathrm{yr}^{-1}$, or $5.7 \%$. 
Since using synthetic line lists is computationally more efficient and is able to simulate spectra of quasars at arbitrary redshifts, we decided to use synthetic line lists in the following analyses.

\subsection{Uncertainty dependences on redshift and resolution}\label{sec:zqso_R}
To evaluate how the uncertainties in the velocity drift measurement depend on the quasar redshift and the spectral resolution of the telescope, we set up mock observations the same way as described in Section \ref{sec:realllvssimll}. We used the synthetic line list and assumed the continuum shape and flux density to be that of S5~0014+81. When analyzing the uncertainty dependence on quasar redshift, we set the resolution to be a fixed value of $R=50,000$, while when analyzing the dependence on resolution we set the quasar redshift to be 3.

\begin{figure*}
    \includegraphics[width=\columnwidth]{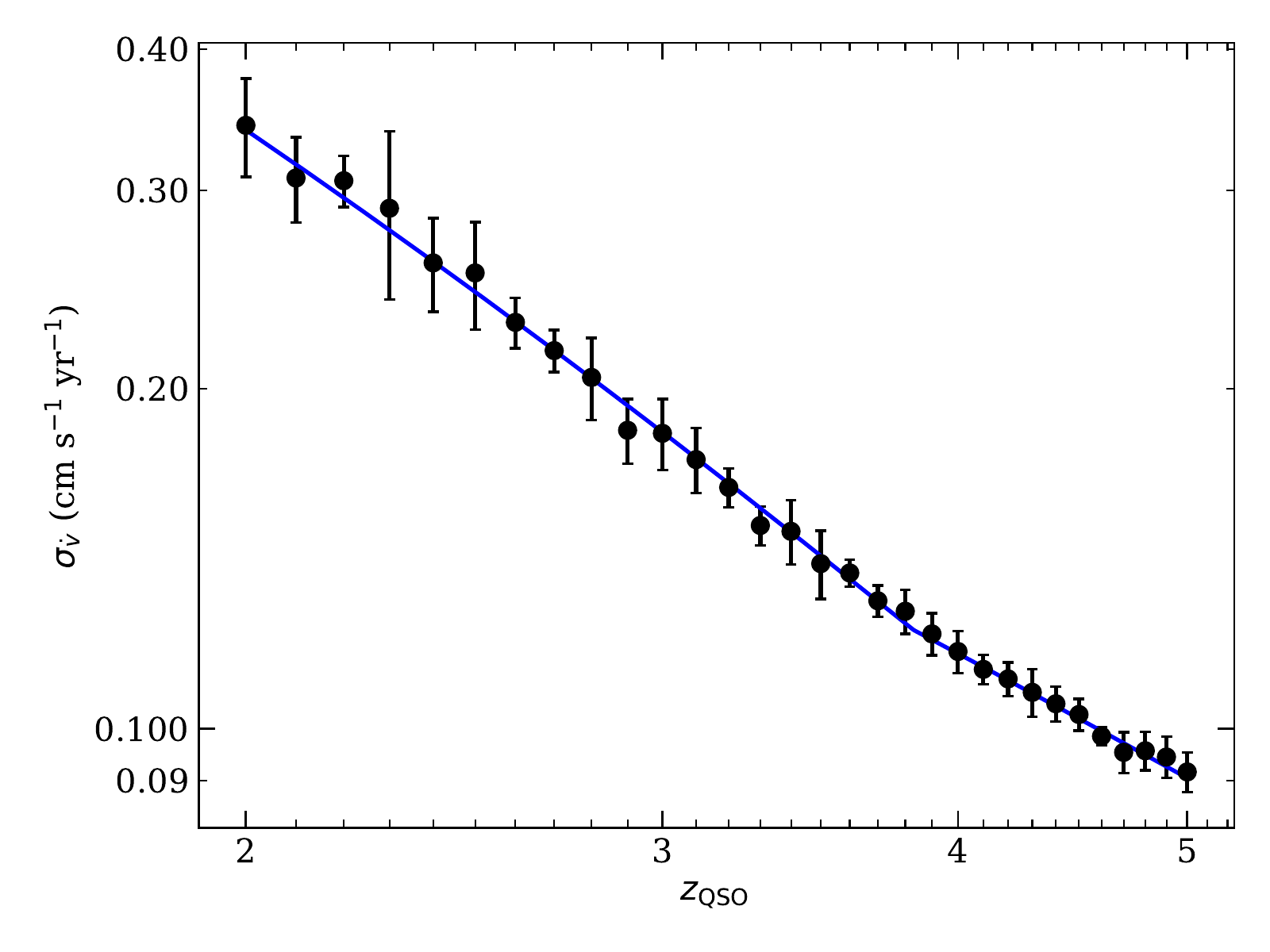}
    \includegraphics[width=\columnwidth]{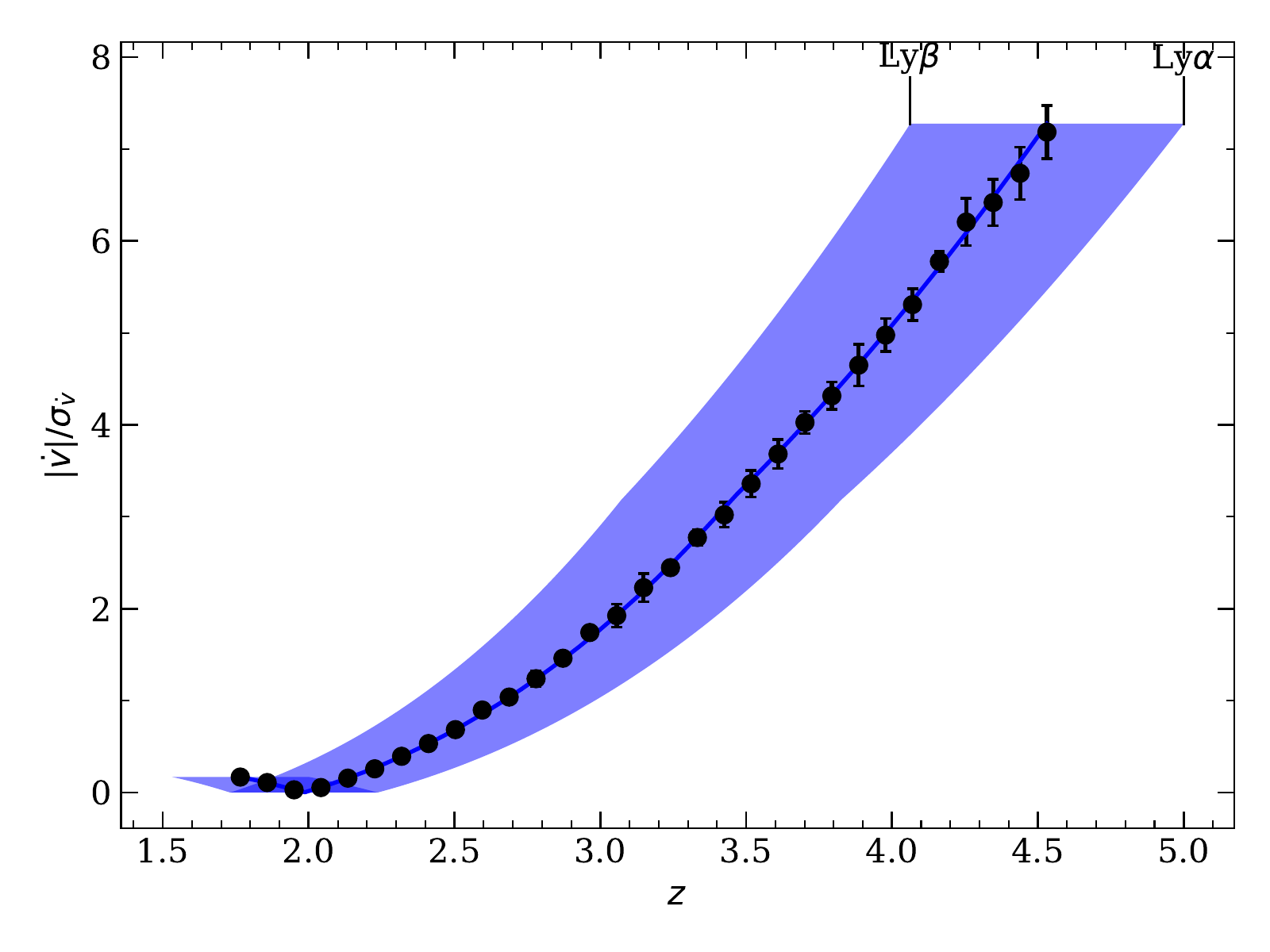}
    \caption{
    The estimated uncertainty dependence on redshift. 
    Left: The expected redshift drift uncertainty as a function of quasar redshift.
    Black points show the estimates and the blue curve is the best-fit relation. Right: $|\dot{v}|/\sigma_{\dot{v}}$ as a function of redshift. The blue shaded region shows the redshift span of the Ly$\alpha$ forest. Black points show the estimates at the redshift midpoints of the Ly$\alpha$ forest and the blue curve is the best-fit relation.
    Each data point is generated assuming the spectral flux density of S5~0014+81, under a fiducial observational setup described in Section \ref{sec:realllvssimll} and fixing the spectral resolution at $R=50,000$. The duration of the experiment is assumed to be 20 years. We note that $\sigma_{\dot{v}}$ scales as $t^{-1/2}$, and so this and other figures can easily be rescaled to other time baselines. The error bars are standard deviations of a sample of 10 randomly generated synthetic line list realizations.
    }\label{fig:sigma_zqso}
\end{figure*}

We generated 31 mock observations of quasars with redshifts from $z=2$ to $z=5$ at $\delta z=0.1$ step size. The $\sigma_{\dot{v}}$ and $|\dot{v}|/\sigma_{\dot{v}}$ dependence on quasar redshifts are shown in Figure~\ref{fig:sigma_zqso}.

We fit a piecewise power law function with two sections to find the $\sigma_{\dot{v}}$ dependence on quasar redshift, while allowing the redshift dividing the two power laws to be a parameter in the fit. The extracted $\sigma_{\dot{v}}$ dependence on quasar redshift is
\begin{equation}
    \sigma_{\dot{v}} \propto (1+z_\mathrm{QSO})^{\alpha},
\end{equation}
where $\alpha = -2.14 \pm 0.04$ for $z_\mathrm{QSO} < z_\mathrm{break}$, and $\alpha = - 1.39 \pm 0.07$ for $z_\mathrm{QSO} > \mathrm{break}$, where $z_\mathrm{break} = 3.83 \pm 0.07$.
The exponent is $\sim0.5$ less than that derived by \cite{Liske08}, 
for which the exponent is $-1.7$ for $z_\mathrm{QSO} < 4$ and $-0.9$ for $z_\mathrm{QSO} > 4$. 
The difference comes from two aspects. First, we are using the continuum shape and flux density of a real quasar, where the S/N is dependent on $z_\mathrm{QSO}$, while \cite{Liske08} were assuming a constant S/N across the spectra. Second, we are assuming a fixed resolution of the spectra, thus the dispersion $\delta \lambda$ is varying across the spectra, while \cite{Liske08} were assuming a fixed dispersion $\delta \lambda=0.0125 \AA$. When fixing $\delta \lambda$, the number of pixels in the Ly$\alpha$ forest is also dependent on $z_\mathrm{QSO}$, such that $N_\mathrm{pix} \propto (1+z_\mathrm{QSO})$. Therefore, the $\sigma_{\dot{v}} \sim z_\mathrm{QSO}$ dependence under \cite{Liske08} assumption can be decomposed as
\begin{equation}
    \sigma_{\dot{v}} (\mathrm{fix\ S/N,\ fix\ } \delta \lambda) \propto N_\mathrm{pix}^{-0.5} (1+z_\mathrm{QSO})^{\alpha_0},
\end{equation}
where $\alpha_0$ is the exponent that only depends on the number of lines. For $z_\mathrm{QSO} < 4$, $\alpha_0 = -1.7 + 0.5 = -1.2$ and for $z_\mathrm{QSO} > 4$, $\alpha_0 = -0.9 + 0.5 = -0.4$, derived by \cite{Liske08}.

On the other hand, under our assumption, the number of pixels is fixed for different $z_\mathrm{QSO}$s because of a fixed resolution. However, the S/N is dependent on $z_\mathrm{QSO}$ in two ways. 
First, at higher redshifts, the energy of a photon decreases because of longer wavelengths, thus the number of photons increases for a given flux density. The dependence scales as $N_\mathrm{phot} \propto \lambda \propto (1+z_\mathrm{QSO})$.
Second, higher dispersion also results in higher $N_\mathrm{phot}$ for a given flux density $F_\lambda$, which scales as $N_\mathrm{phot} \propto \delta \lambda \propto (1+z_\mathrm{QSO})$.
Combining the two factors, the $\mathrm{S/N} \propto N_\mathrm{phot}^{0.5} \propto (1+z_\mathrm{QSO})$. Therefore, the $\sigma_{\dot{v}} \sim z_\mathrm{QSO}$ dependence under our assumption can be decomposed as
\begin{equation}
    \sigma_{\dot{v}} (\mathrm{fix\ } F_\lambda, \mathrm{fix\ } R) \propto \mathrm{S/N}^{-1} (1+z_\mathrm{QSO})^{\alpha_0}.
\end{equation}
We derive that for $z_\mathrm{QSO} < z_\mathrm{break}$, $\alpha_0 = -2.14 + 1 = -1.14$ and for $z_\mathrm{QSO} > z_\mathrm{break}$, $\alpha_0 = -1.39 + 1 = -0.39$. Therefore, the $\alpha_0$ exponent dereived by us is consistent with \cite{Liske08}, while the $\alpha$ exponent is different by $\sim 0.5$.

We generalize the dependence of $\sigma_{\dot{v}}$ on a single quasar $i$ for our analysis as
\begin{equation}
\begin{aligned}
\sigma_{\dot{v}_i} = (0.12 \pm 0.01)
\left(\frac{\overline{F_{\lambda, i}}}{10^{-15}\ \mathrm{erg}\ \mathrm{cm}^{-2}\ \mathrm{s}^{-1}\ \mathrm{\AA}^{-1}}\right)^{-1/2}
\left(\frac{1 + z_{\mathrm{QSO}, i}}{1 + z_\mathrm{break}}\right)^{\alpha}&\\
~\mathrm{cm~s}^{-1}\mathrm{~yr}^{-1},&
\end{aligned}
\end{equation}
where $\sigma_{\dot{v}_i}$, $\overline{F_{\lambda, i}}$ and $z_{\mathrm{QSO}, i}$ are for individual quasars, $\overline{F_{\lambda, i}}$ is the mean flux density of the Ly$\alpha$ forest. 
The composite $\sigma_{\dot{v}}$ for an ensemble of $N_{\mathrm{QSO}}$ quasars will then be given by
\begin{equation}
\sigma_{\dot{v}} = \left(\sum_{i=1}^{N_{\mathrm{QSO}}} \sigma_{\dot{v}_i}^{-2}\right)^{-1/2}.
\end{equation}

\begin{figure*}
    \includegraphics[width=\columnwidth]{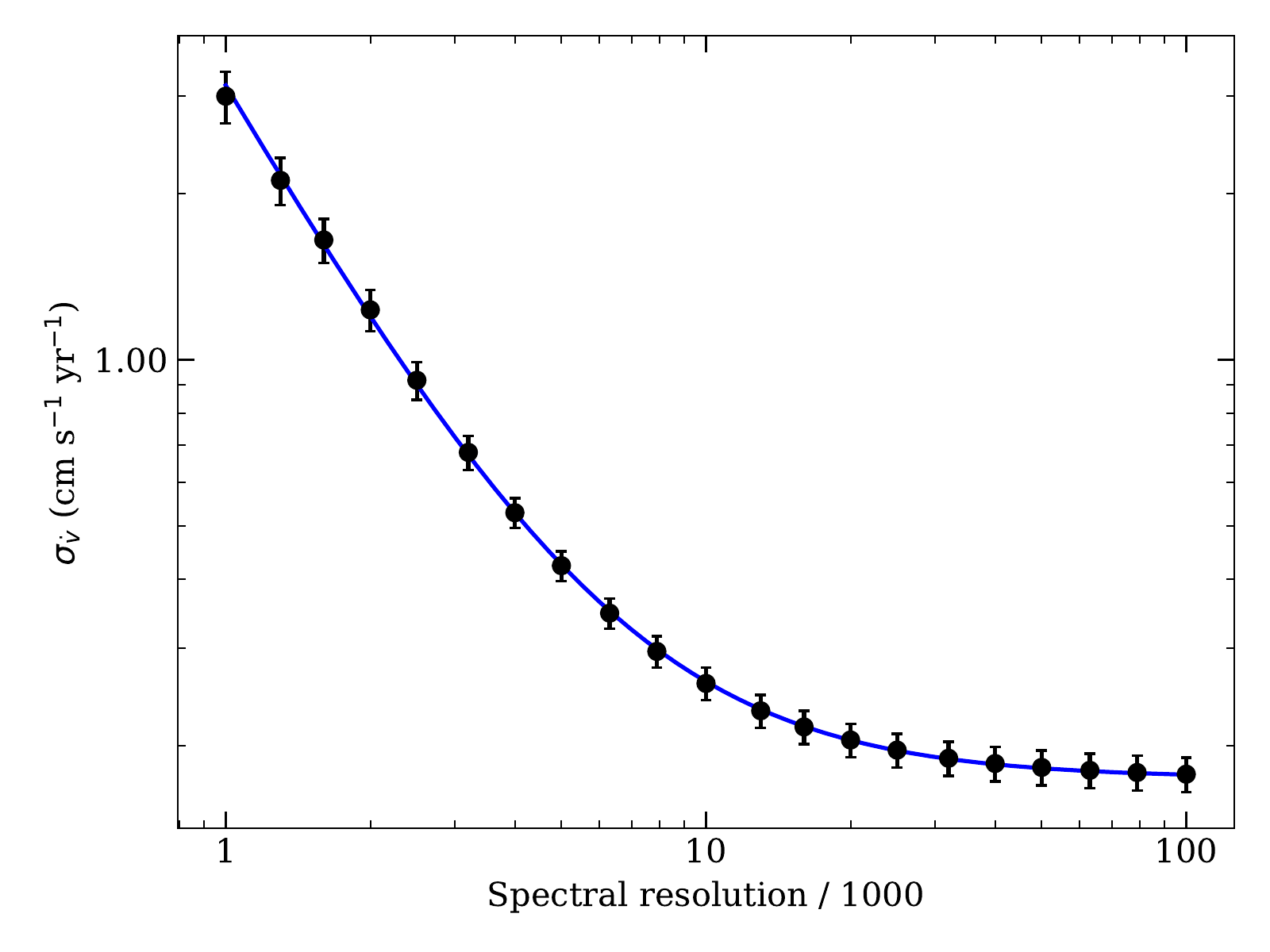}
    \includegraphics[width=\columnwidth]{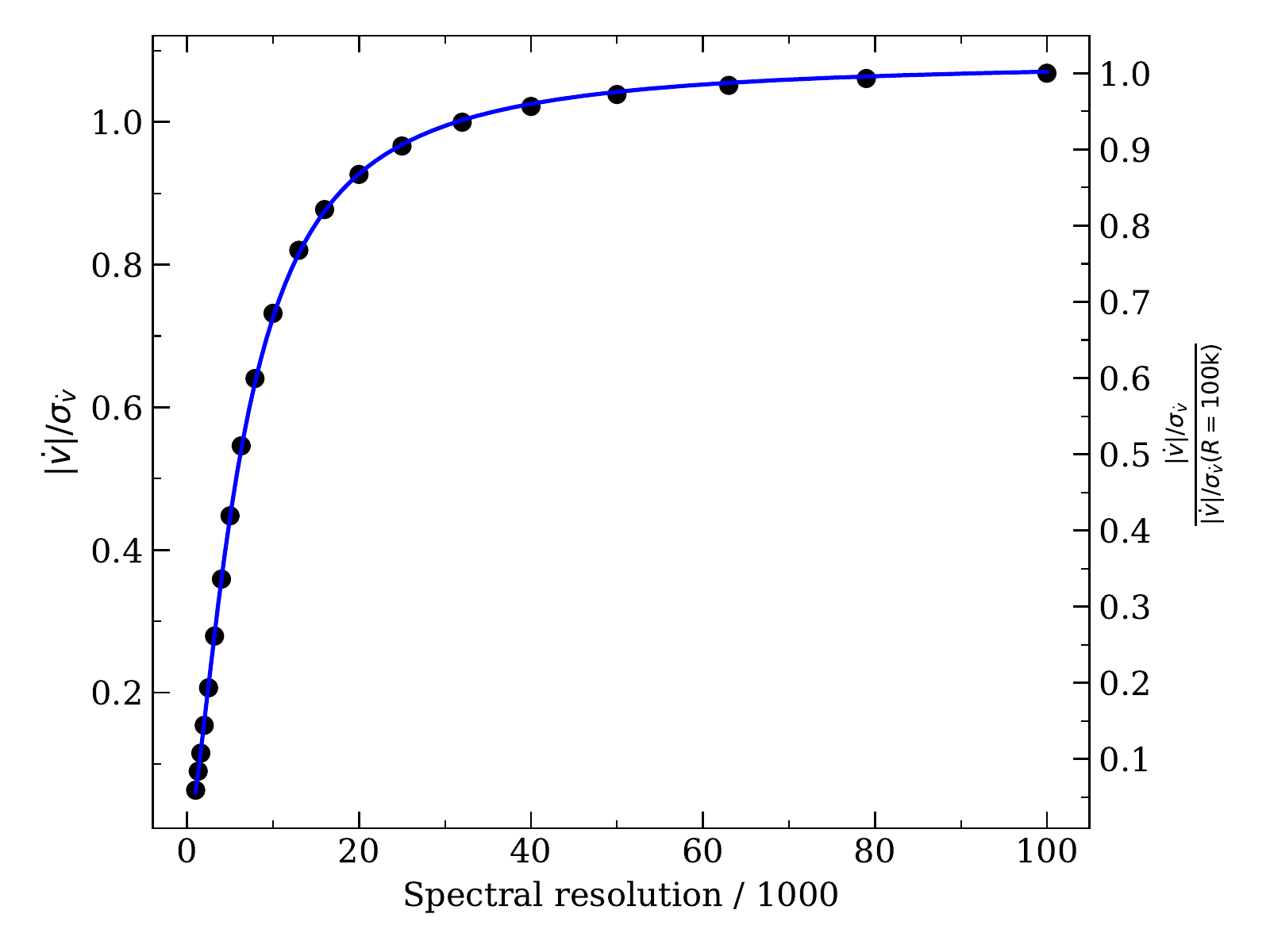}
    \caption{
    The estimated uncertainty dependence on spectral resolution. 
    Left: $\sigma_{\dot{v}}$ as a function of resolution.  Right: $|\dot{v}|/\sigma_{\dot{v}}$ as a function of resolution. For both panels, black points show the estimates and the blue curve is the best-fit relation.
    The observational assumptions are the same as described in Figure~\ref{fig:sigma_zqso} except that the quasar redshift is fixed at $z=3$.
    The duration of the experiment is assumed to be 20 years.
    The resolution improvement stops at $R \sim 20,000$, around the spectral resolution that matches the typical line widths of the Lyman-$\alpha$ forest, because there is limited gain once we resolve the line.
    }\label{fig:sigma_R}
\end{figure*}

We generated 21 mock observations at resolutions from $R=1,000$ to $R=100,000$ with logarithm step sizes.
The $\sigma_{\dot{v}}$ and $|\dot{v}|/\sigma_{\dot{v}}$ dependence on resolution are shown in Figure~\ref{fig:sigma_R}.
The uncertainty drops dramatically as the resolution increases until $R\sim20,000$, after which the $\sigma_{\dot{v}}$ remains nearly constant as the resolution exceeds the typical line width of a Ly$\alpha$ line, as was found by \cite{Liske08}.
The change is precisely because there is limited gain once we resolve the line.
A power law plus a constant bias fits the $\sigma_{\dot{v}}$ dependence on resolution well. The best-fit relation for our particular setup is
\begin{equation}\label{eq:sigma_R}
\begin{aligned}
\sigma_{\dot{v}} = 
(7.3\pm 1.0)
\times 10^{-3} \left(\frac{R}{50,000}\right)^{-1.53 \pm 0.01} &\\
+ (0.1747 \pm 0.0009)
~\mathrm{cm~s}^{-1}\mathrm{~yr}^{-1}.&
\end{aligned}
\end{equation}
Note that this relation is specific to the observation cadence and experiment duration we have chosen --- observing each quasar for 50 hours per week for 20 years.

\subsection{Time allocation}\label{sec:time}
\begin{figure*}
    \includegraphics[width=\columnwidth]{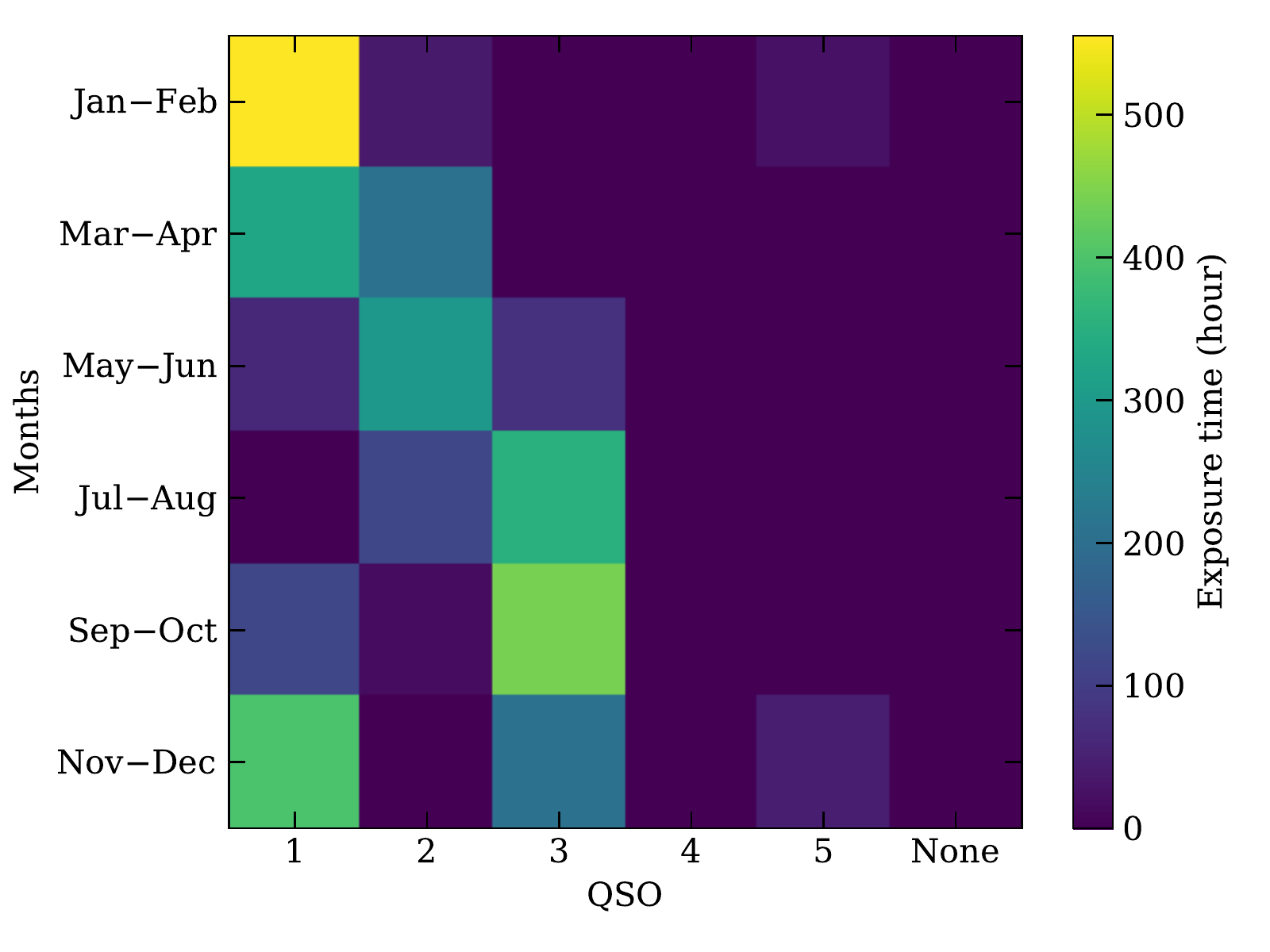}
    \includegraphics[width=\columnwidth]{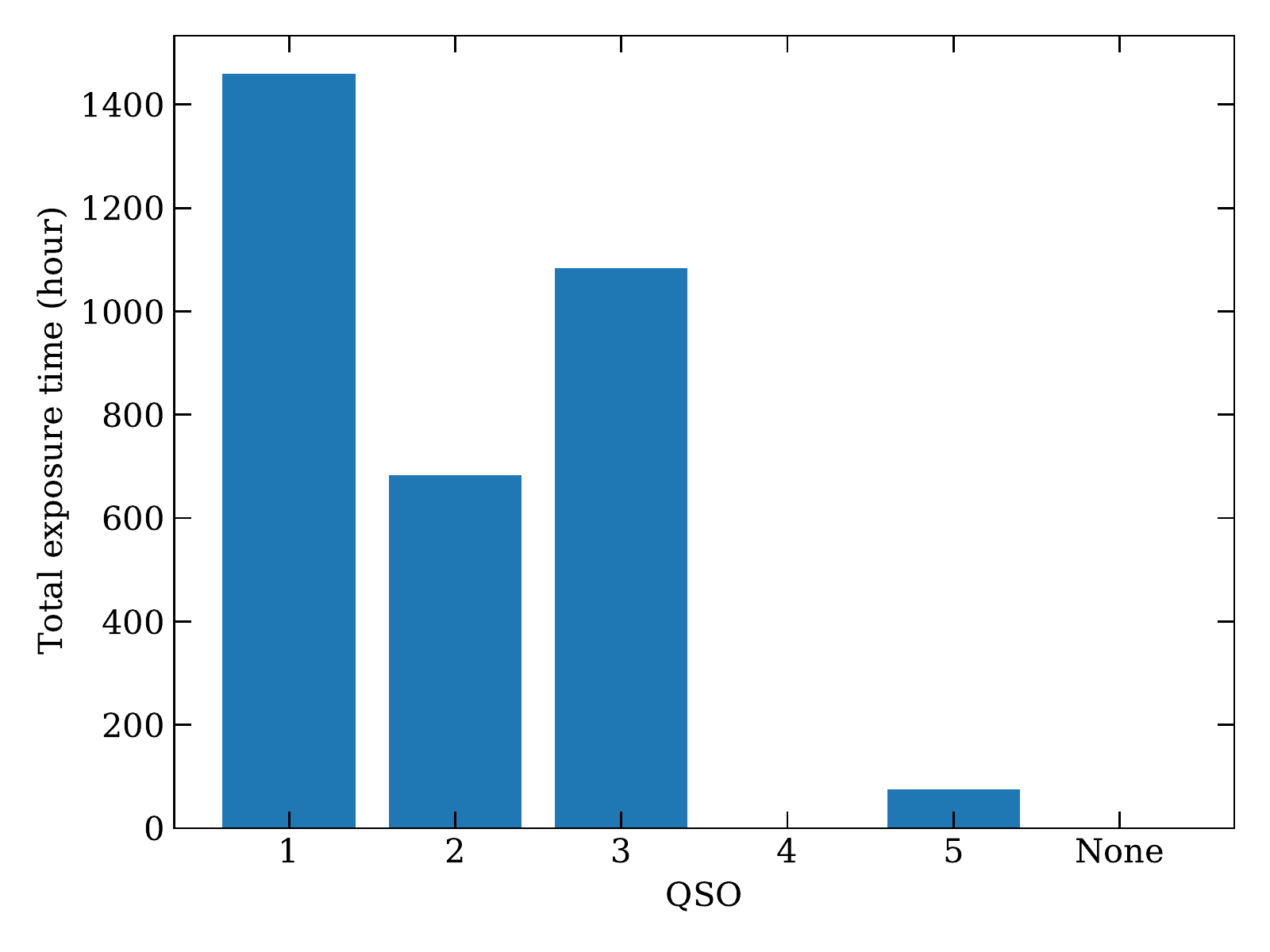}
    \caption{Annual exposure time allocation for a planned experiment at the Roque de los Muchachos Observatory. Left: color coded is the assigned time for each QSO for six 2-month time windows. Right: annual total exposure time for each QSO. For both panels, the QSO numbers correspond to the rank of the northern sample with respect to the expected $|\dot{v}|/\sigma_{\dot{v}}$ (as listed in the first column of Table~\ref{tab:qsos}), i.e. the observing priority. ``None'' corresponds to time when none of these 5 QSOs is observable. 
    This Figure indicates that for the northern sample only 4 quasars (numbers 1, 2, 3, and 5) are sufficient to utilize all the available observing time during a given year.
    }\label{fig:time}
\end{figure*}

Because of the distribution of $|\dot{v}|/\sigma_{\dot{v}}$ values for the targets, the final S/N for this experiment is sensitive to how the time is allocated among the targets. 
In addition to considering a simple uniform time allocation among the proposed quasars, we also consider an observing strategy which prioritizes the highest S/N targets, taking into consideration the on-sky position of the quasars and the observing time over a year. Utilizing the {\scriptsize\texttt{ASTROPLAN}} package \citep{Morris18}, we plan the observations for each night during the year 2021, and integrate the exposure time for each target. 
Here we consider observations of the northern hemisphere sample. Similar results would be obtained from Hawaii, Arizona, or the Canary Islands; for concreteness we set the observing site to be the Roque de los Muchachos Observatory ($28^\circ45'30''$ N, $17^\circ52'48''$ W) in our calculations.
We also introduce the observation plan of the southern hemisphere sample in Section \ref{sec:southern} in the appendix.

The observing strategy is as follows. We give higher priority to quasars with higher predicted $|\dot{v}|/\sigma_{\dot{v}}$. Starting from the astronomical twilight evening of Jan 1st, we begin with the quasar with the highest priority. If that quasar is $>30^\circ$ above horizon (with airmass $<2$) at that time, we choose to observe that quasar, until it falls below $30^\circ$ above horizon, or a higher priority quasar rises above $30^\circ$ above horizon, or it reaches the astronomical twilight morning. If the highest priority quasar is less than $30^\circ$ above horizon, we check the quasar with the next highest priority, repeating until an observable target is found. At the end of an observation of a quasar, we check again for the highest priority quasar. We follow this strategy every night until the end of the year, always observing the highest $|\dot{v}|/\sigma_{\dot{v}}$ quasar visible at least $30^\circ$ above the horizon.

Figure~\ref{fig:time} shows the integrated time allocation for each quasar over six 2-month windows of the year. The quasars are numbered by their observing priorities. The first quasar (APM~08279+5255) is allocated most time in winter, with 1460.2 hours of total integration time over the year. The second (B1422+231) and third (PS1~J212540.96-171951.4) quasar are allocated most time in spring and fall, with 682.3 and 1083.0 hours of total integration time, respectively. The fourth quasar (PSS~J0926+3055) is not allocated time because it is close to the first quasar, which has higher priority. The fifth quasar (SDSS~J034151.16+172049.7) is allocated a small amount of time (75.4 hours) and there are no time left after assigning time to the top 5 quasars. The overall total integration time of the year is 3300.9 hours. We discuss the simulation results of this observing plan in Section \ref{sec:prospect}.

\subsection{Prospects for a detection}\label{sec:prospect}
Using our $|\dot{v}|/\sigma_{\dot{v}}$ selected samples, we evaluate the observational setup required to obtain a $|\dot{v}|/\sigma_{\dot{v}} = 3$ ($3\sigma$ detection) and $|\dot{v}|/\sigma_{\dot{v}} = 5$ ($5\sigma$ detection) redshift drift detection. 
We adopt a fiducial resolution $R=50,000$ as being approximately mid-range for plausible instrumental designs. The gain from going to higher resolution is minimal, as the expected S/N at R = 50,000 is 97\% of that at $R = 100,000$ (Figure~\ref{fig:sigma_R}). Going to lower resolution may be desirable for minimizing cost, as it is possible to decrease to $R=20,000$ with only modest loss (S/N 87\% of that at $R=100,000$). However, for $R<20,000$  the uncertainty rises rapidly (Figure~\ref{fig:sigma_R}).
We do note that other high precision radial velocity experiments for stellar astrophysics or exoplanet searches require higher spectral resolution ($R=100,000$). Such considerations may drive one to higher resolution, as may technical issues. Doing so has no negative impact on a redshift drift experiment other than increased cost.
A total efficiency of $\epsilon = 25\%$ is assumed. 

We know that $|\dot{v}|/\sigma_{\dot{v}}$ is proportional to the aperture $D$, and is proportional to $t^{3/2}$ because $|\dot{v}|\sim t$ and $\sigma_{\dot{v}}\sim t^{-1/2}$. We simulate observations for a 15~m telescope with a time baseline of 20 years to calculate the $|\dot{v}|/\sigma_{\dot{v}}$ for three different observational strategies.
We then plot lines of constant $|\dot{v}|/\sigma_{\dot{v}}$ for $|\dot{v}|/\sigma_{\dot{v}}=3$ and $|\dot{v}|/\sigma_{\dot{v}}=5$ as a function of aperture and time baseline.

The first strategy is to observe the most optimal target all the time, and the target is on sky for all nights, which is the idealized situation. The second strategy is to observe the 5 most optimal targets and uniformly distribute the observing time into the five. The third strategy is to observe the 5 most optimal targets following the time allocation strategy as described in Section \ref{sec:time}, which is a realistic situation.

\begin{figure}
    \includegraphics[width=\columnwidth]{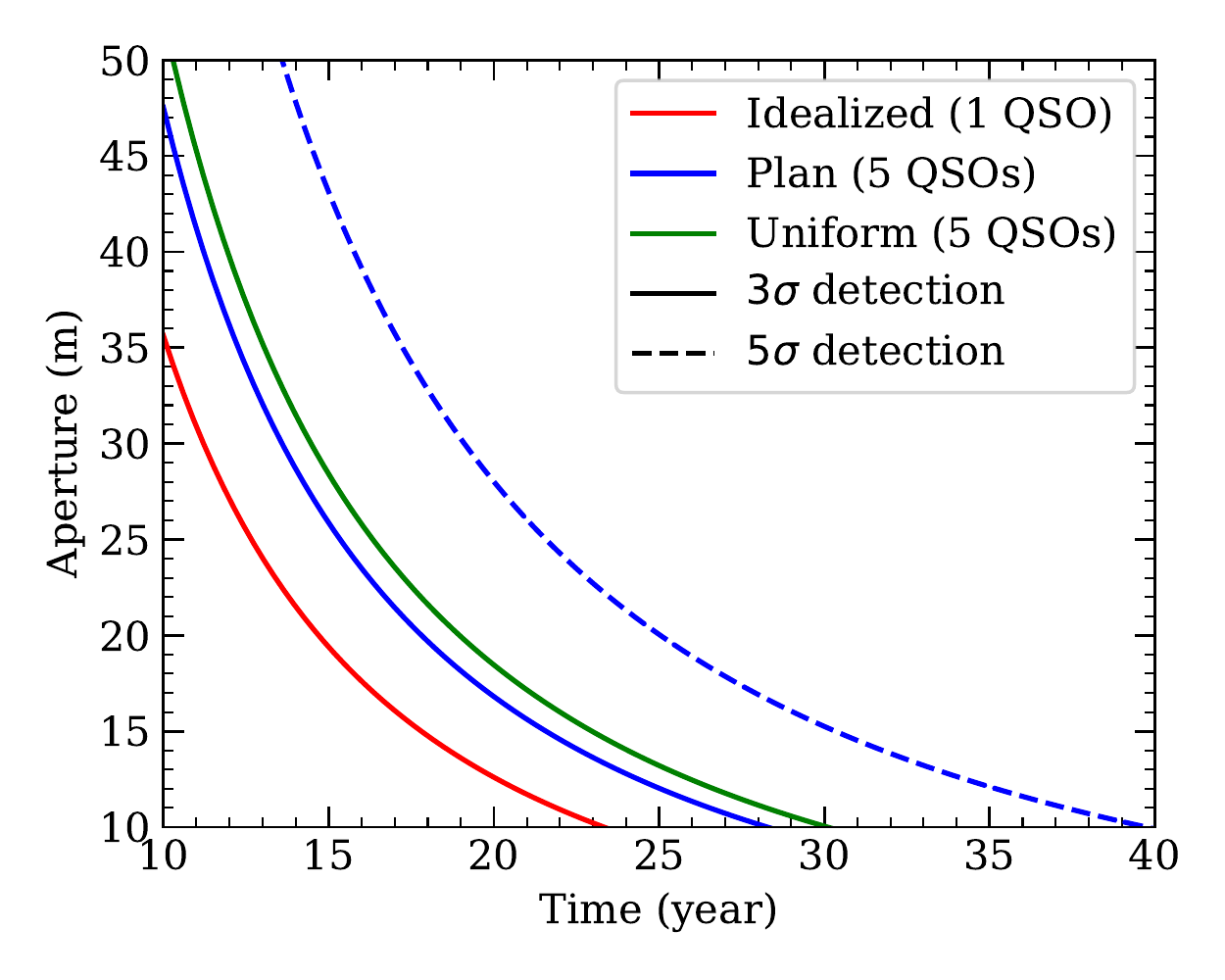}
    \caption{
    The telescope aperture and observing time required for a $|\dot{v}|/\sigma_{\dot{v}} = 3$ ($3\sigma$ detection; solid curve) and $|\dot{v}|/\sigma_{\dot{v}} = 5$ ($5\sigma$ detection; dashed curve), color-coded by the strategy of time allocation among quasars. The legends stating ``Idealized (1~QSO)'' (red curve) or ``Uniform (5~QSOs)'' (green curve) indicate observing only the best (highest predicted $|\dot{v}|/\sigma_{\dot{v}}$) quasar or applying uniform time distribution among the best 5 quasars, respectively.
    It is not in practice possible to observe only the best quasar all the time, which is why we are stating this as the idealized case.
    The line labeled with ``Plan (5~QSOs)'' (blue curve) indicates the realistic time allocation plan described in Section \ref{sec:time}.
    All time allocation strategies have the same overall total exposure time.
    }\label{fig:aperture_period}
\end{figure}

For the first and second strategies, we consider mock observations that are scheduled 50 hours per week with uniform time spent on each quasar and one composite spectrum generated every two months, resulting in 2640 hours overall total exposure time per year. We tested observing the top 1, 2, 3, 5 and 10 quasars in term of the predicted $|\dot{v}|/\sigma_{\dot{v}}$ during the experiment with equal time allocation among the quasars. The result indicates that dedicating time to few best quasars gives better redshift drift detection than distributing time to a larger number of not-as-good quasars.
The caveat is that with only a few quasars, the sensitivity of the experiment may be adversely affected if one of the targets dims significantly during the period of the experiment.
The resulting aperture and time requirements using 1 and 5 quasars are shown as the red and green curves in Figure~\ref{fig:aperture_period}, corresponding to the first and second observing strategies, respectively. 

For the third strategy, we adopt the non-uniform time allocation strategy described in Section \ref{sec:time}, which is designed to mimic a realistic observing strategy. We apply a weather condition and miscellaneous time loss of 20\% in the observation plan, resulting in an overall total exposure time of 2640.7 hours, which is consistent with the uniform time allocation plan. The best-fit relation using this strategy is 
\begin{equation}\label{eq:aperture_time}
|\dot{v}|/\sigma_{\dot{v}} = 3.57 
\left(\frac{D}{20\ \text{m}}\right)
\left(\frac{t}{20\ \text{yr}}\right)^{3/2}
\left(\frac{\epsilon}{0.25}\right)^{1/2},
\end{equation}
which is again under \cite{Planck16} cosmology. 
For other cosmological models, it is straightforward to just change the $\dot{v}$ expression to derive this relation.
The $|\dot{v}|/\sigma_{\dot{v}}$ is also related to the spectral resolution $R$ through $\sigma_{\dot{v}}$, as discussed in Section \ref{sec:zqso_R}.
The corresponding curve of Equation \ref{eq:aperture_time} is shown as the blue curve in Figure~\ref{fig:aperture_period}, corresponding to the third observing strategy.
Using our time allocation plan, the $|\dot{v}|/\sigma_{\dot{v}}$ for a given aperture and time is better than uniformly allocating time to the best 5 quasars, but is not as good as only observing the best one quasar. Further tests indicate that the $|\dot{v}|/\sigma_{\dot{v}}$ for the plan is equivalent to uniformly allocating time to the best $\sim 3$ quasars in the idealized case where those 3 quasars can be observed throughout the year.

The simulation results of \cite{Liske08} showed that for a 42~m telescope, dedicating 200 hours per year for 20 years could yield a $3\sigma$ detection using 10 QSOs, or $4\sigma$ detection if instead targeting the 3 highest $|\dot{v}|/\sigma_{\dot{v}}$ QSOs, demonstrating that the greatest leverage comes from focusing on the most optimal targets.
For the same aperture telescope, the situation is improved given the new, brighter quasar sample in this work. Using our dedicated observing strategy and observing the 5 highest $|\dot{v}|/\sigma_{\dot{v}}$ QSOs, one could achieve a $3\sigma$ detection within 10.9 years or a $4\sigma$ detection within 13.2 years.
Additional brighter quasars would further decrease the required time for a detection.

\begin{figure*}
    \includegraphics[width=\columnwidth]{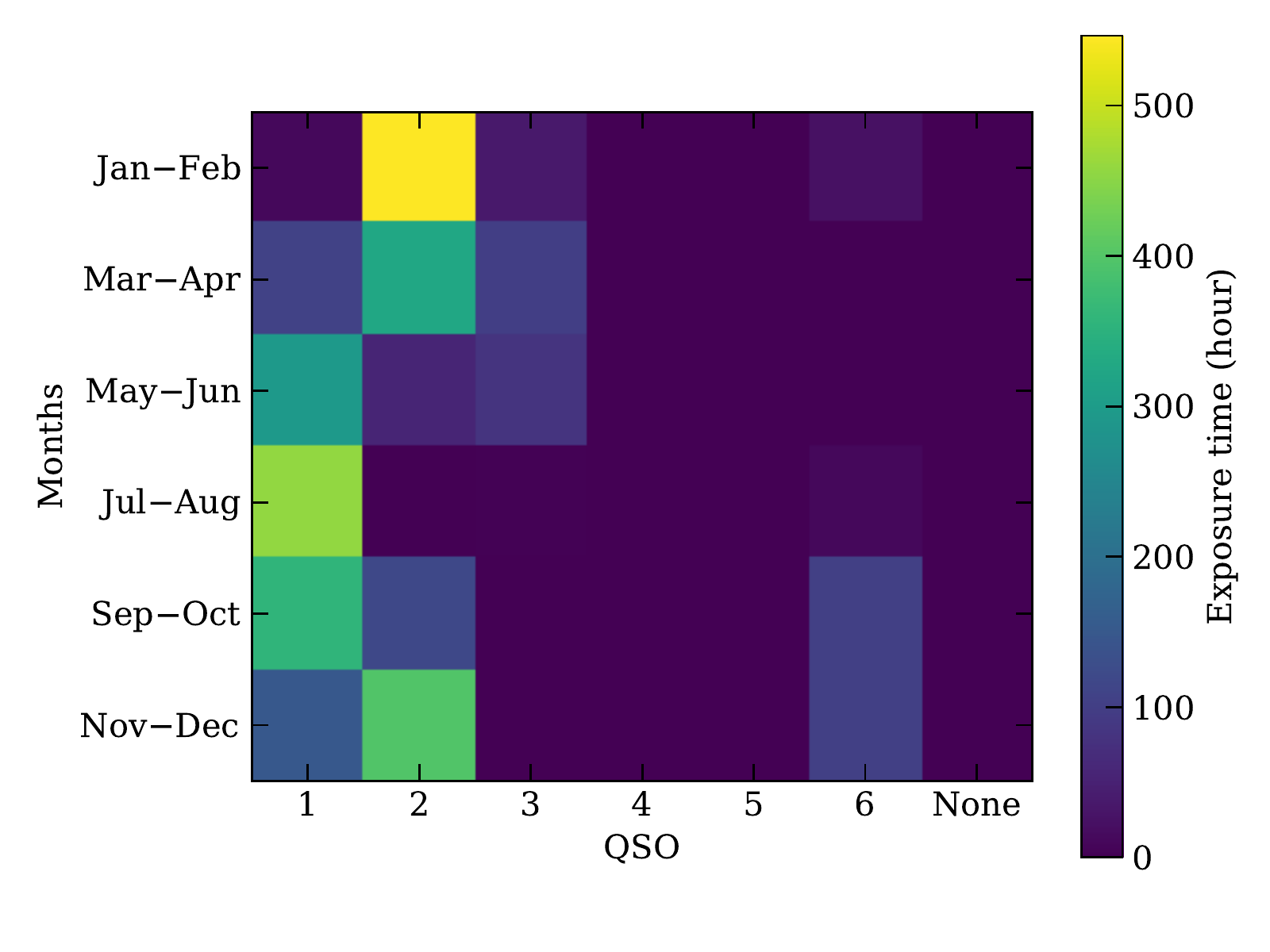}
    \includegraphics[width=\columnwidth]{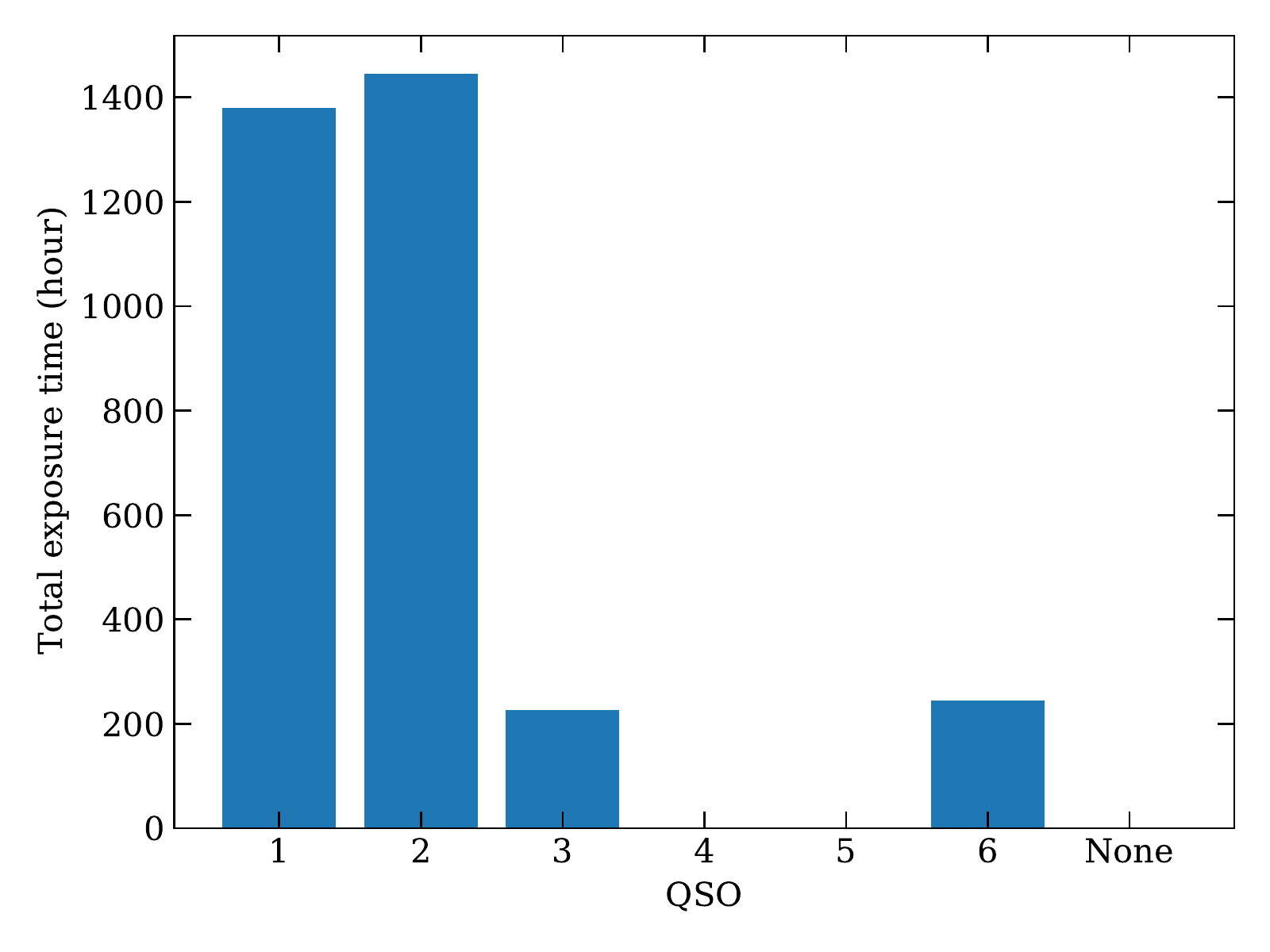}
    \caption{Annual exposure time allocation, same as Figure~\ref{fig:time}, but now assuming we find a new quasar that is as bright as or brighter than the current best quasar, and is located on the opposite side to the current best quasar on the sky. In this figure, the new quasar is labeled as number 1 and the previous number 1--5 quasars in Figure~\ref{fig:time} are now labeled number 2--6. This figure indicates that if we are able to find a new quasar as described here, most time will be allocated to the best 2 quasars, thus improve the efficiency of the redshift drift experiment.
    }\label{fig:time_newqso}
\end{figure*}

\begin{figure}
    \includegraphics[width=\columnwidth]{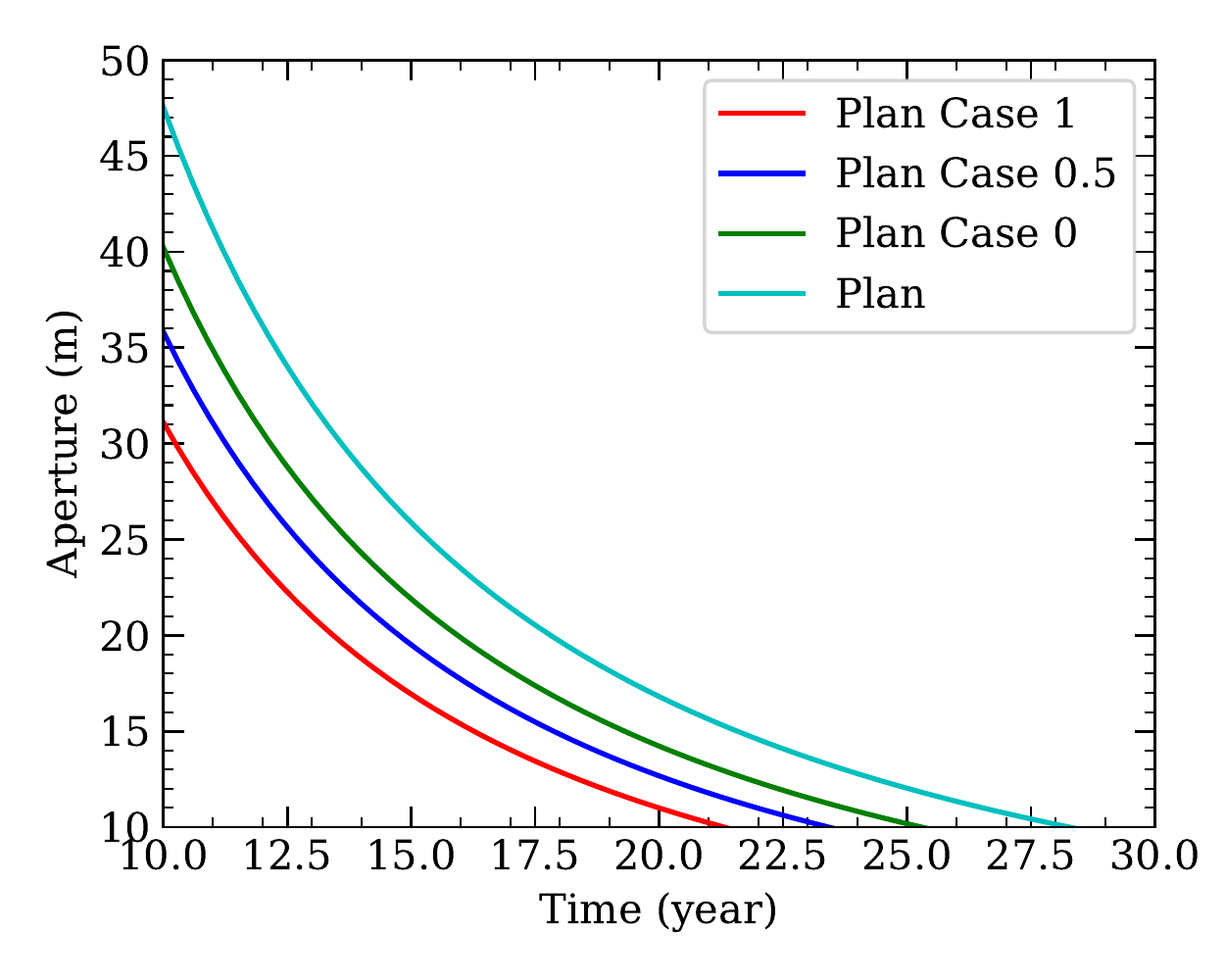}
    \caption{
    Similar to Figure~\ref{fig:aperture_period}, shown here are the telescope aperture and observing time required for a $|\dot{v}|/\sigma_{\dot{v}} = 3$ ($3\sigma$ detection), using our optimal time allocation plan described in Section \ref{sec:time}.
    We compare scenarios where we use currently discovered quasars (labeled ``Plan'' and drawn as the cyan curve) and where we are able to find a new quasar that is as bright as or brighter than the current best quasar (labeled ``Plan Case $x$'', where $x = 0, 0.5, 1$ drawn as green, blue and red curves indicate the new quasar is 0, 0.5 and 1 magnitude brighter than the current best quasar, respectively).
 }\label{fig:aperture_period_addqso}
\end{figure}

There are attempts to find new luminous quasars to improve the performance of the redshift drift experiment \citep[e.g.,][]{Schindler17,Schindler19b,Boutsia20,Jeram20,Wolf20}. Here we also tested scenarios where we find a new quasar visible in the northern hemisphere that is 0, 0.5, or 1 magnitude brighter than the current best quasar, and is on the opposite side to the current best quasar on the sky (having an R.A. 180~degrees different from the current best quasar, but having the same declination).
Placing the new quasar on the opposite side of the sky is an optimistic scenario, which we use to illustrate the maximal gain that might be achieved through discovery of a new, bright quasar.
The time allocation using this strategy is shown in Figure~\ref{fig:time_newqso}. In the figure, the new quasar is labeled as number 1 and the previous number 1--5 quasars are labeled number 2--6 now. By adding the new quasar on the opposite side to the current best quasar, most time will be allocated to the best 2 quasars. 

To estimate the aperture and time requirements for the new strategy with the added quasar, we do a simple test by changing the $M_{1450}$ of the current best quasar while leaving other properties of the quasar to simulate the new quasar. The aperture and time requirements for scenarios we find a 0, 0.5, or 1 magnitude brighter quasar are shown as the green, blue and red lines in Figure~\ref{fig:aperture_period_addqso}, respectively. We find that adding a new quasar with the same brightness as the current best quasar can increase the S/N by 18\%, or equivalently save us 11\% of the time.
The situation improves as brighter quasars are found. For example, if we are able to find a quasar that is 1 magnitude brighter than the current best quasar, with only a 20~m telescope we can achieve a 3$\sigma$ detection in 13.4~years.

\section{Conclusions}\label{sec:conclusion}
We discussed different aspects of planning a redshift drift experiment, and performed end-to-end simulations on the experiment to explore the observational requirements for a success redshift drift detection.
Relative to \cite{Liske08}, two main refinements in the current work are inclusion of quasars from more recent catalogs and consideration of a realistic observing strategy for a dedicated redshift drift experiment that maximizes $\dot{v}/\sigma_{\dot{v}}$.
\begin{enumerate}
\item We have developed an end-to-end code for simulating Lyman-$\alpha$ forest redshift drift observations.
In developing this code, we demonstrate that synthetic spectral templates yield consistent redshift drift results to reconstructed HIRES spectral templates from the Keck archive.

\item We define samples of the best targets based upon $|\dot{v}|/\sigma_{\dot{v}}$ and test observing strategies. From designing a realistic mock observing plan, we find that an observing strategy that optimizes S/N includes observations of just four targets for a facility in the northern hemisphere. We also find that allocating the observing time to target at any time the visible quasar with the highest $|\dot{v}|/\sigma_{\dot{v}}$ yields 10\% higher SNR than if one were to have uniformly observed the 5 highest $|\dot{v}|/\sigma_{\dot{v}}$ quasars.

\item In considering the sensitivity of a dedicated facility, we find that redshift drift can be detected via Lyman-$\alpha$ forest observation at $\mathrm{S/N}=3~(5)$ in 15~(22) years with a 25~m telescope, given a spectrograph with long term stability with $R=50,000$ and an efficiency of 25\%. Any efficiency gain will also shorten this time; if the efficiency is instead 35\%, then the time to achieve $\mathrm{S/N}=3$ decreases to 13.7 years.
\end{enumerate}

The aim of this paper has been to assess the feasibility of obtaining a first detection of redshift drift with a dedicated facility.  
Our results indicate that given an observatory such as the one described in \cite{Eikenberry19a,Eikenberry19b,Eikenberry19c}, such a detection is achievable within 15 years.
With the attempts of finding new luminous quasars, the expected detection can be achieved with smaller telescopes or shorter time.

\section*{Acknowledgements}
This research has made use of the Keck Observatory Archive (KOA), which is operated by the W. M. Keck Observatory and the NASA Exoplanet Science Institute (NExScI), under contract with the National Aeronautics and Space Administration.

\section*{Data Availability}
The Keck and SDSS data can be accessed from their online archives. The Keck Observatory Archive website is https://www2.keck.hawaii.edu/koa/public/koa.php. The SDSS website is https://www.sdss.org.

\bibliographystyle{mnras}
\bibliography{main} 

\begin{thebibliography}{}
\makeatletter
\relax
\def\mn@urlcharsother{\let\do\@makeother \do\$\do\&\do\#\do\^\do\_\do\%\do\~}
\def\mn@doi{\begingroup\mn@urlcharsother \@ifnextchar [ {\mn@doi@}
  {\mn@doi@[]}}
\def\mn@doi@[#1]#2{\def\@tempa{#1}\ifx\@tempa\@empty \href
  {http://dx.doi.org/#2} {doi:#2}\else \href {http://dx.doi.org/#2} {#1}\fi
  \endgroup}
\def\mn@eprint#1#2{\mn@eprint@#1:#2::\@nil}
\def\mn@eprint@arXiv#1{\href {http://arxiv.org/abs/#1} {{\tt arXiv:#1}}}
\def\mn@eprint@dblp#1{\href {http://dblp.uni-trier.de/rec/bibtex/#1.xml}
  {dblp:#1}}
\def\mn@eprint@#1:#2:#3:#4\@nil{\def\@tempa {#1}\def\@tempb {#2}\def\@tempc
  {#3}\ifx \@tempc \@empty \let \@tempc \@tempb \let \@tempb \@tempa \fi \ifx
  \@tempb \@empty \def\@tempb {arXiv}\fi \@ifundefined
  {mn@eprint@\@tempb}{\@tempb:\@tempc}{\expandafter \expandafter \csname
  mn@eprint@\@tempb\endcsname \expandafter{\@tempc}}}

\bibitem[\protect\citeauthoryear{{Akaike}}{{Akaike}}{1974}]{Akaike74}
{Akaike} H.,  1974, IEEE Transactions on Automatic Control, 19, 716

\bibitem[\protect\citeauthoryear{{Bautista} et~al.,}{{Bautista}
  et~al.}{2017}]{Bautista17}
{Bautista} J.~E.,  et~al., 2017, \mn@doi [\aap] {10.1051/0004-6361/201730533},
  \href {https://ui.adsabs.harvard.edu/abs/2017A&A...603A..12B} {603, A12}

\bibitem[\protect\citeauthoryear{{Berriman}, {Ciardi}, {Laity}, {Tahir-Kheli},
  {Conrad}, {Mader}, {Tran}  \& {Bida}}{{Berriman} et~al.}{2005}]{Berriman05}
{Berriman} G.~B.,  {Ciardi} D.~R.,  {Laity} A.~C.,  {Tahir-Kheli} N.~D.,
  {Conrad} A.,  {Mader} J.,  {Tran} H.,   {Bida} T.,  2005, in {Shopbell} P.,
  {Britton} M.,   {Ebert} R.,  eds,  Astronomical Society of the Pacific
  Conference Series Vol. 347, Astronomical Data Analysis Software and Systems
  XIV. p.~627

\bibitem[\protect\citeauthoryear{{Berriman} et~al.,}{{Berriman}
  et~al.}{2014}]{Berriman14}
{Berriman} G.~B.,  et~al., 2014, in {Chiozzi} G.,  {Radziwill} N.~M.,  eds,
  Society of Photo-Optical Instrumentation Engineers (SPIE) Conference Series
  Vol. 9152, Software and Cyberinfrastructure for Astronomy III. p. 91520A
  (\mn@eprint {arXiv} {1408.0835}), \mn@doi{10.1117/12.2054772}

\bibitem[\protect\citeauthoryear{{Blanton} et~al.,}{{Blanton}
  et~al.}{2017}]{Blanton17}
{Blanton} M.~R.,  et~al., 2017, \mn@doi [\aj] {10.3847/1538-3881/aa7567}, \href
  {https://ui.adsabs.harvard.edu/abs/2017AJ....154...28B} {154, 28}

\bibitem[\protect\citeauthoryear{{Bouchy}, {Pepe}  \& {Queloz}}{{Bouchy}
  et~al.}{2001}]{Bouchy01}
{Bouchy} F.,  {Pepe} F.,   {Queloz} D.,  2001, \mn@doi [\aap]
  {10.1051/0004-6361:20010730}, \href
  {https://ui.adsabs.harvard.edu/abs/2001A&A...374..733B} {374, 733}

\bibitem[\protect\citeauthoryear{{Boutsia} et~al.,}{{Boutsia}
  et~al.}{2020}]{Boutsia20}
{Boutsia} K.,  et~al., 2020, \mn@doi [\apjs] {10.3847/1538-4365/abafc1}, \href
  {https://ui.adsabs.harvard.edu/abs/2020ApJS..250...26B} {250, 26}

\bibitem[\protect\citeauthoryear{{Calderone} et~al.,}{{Calderone}
  et~al.}{2019}]{Calderone19}
{Calderone} G.,  et~al., 2019, \mn@doi [\apj] {10.3847/1538-4357/ab510a}, \href
  {https://ui.adsabs.harvard.edu/abs/2019ApJ...887..268C} {887, 268}

\bibitem[\protect\citeauthoryear{{Darling}}{{Darling}}{2012}]{Darling12}
{Darling} J.,  2012, \mn@doi [\apjl] {10.1088/2041-8205/761/2/L26}, \href
  {https://ui.adsabs.harvard.edu/abs/2012ApJ...761L..26D} {761, L26}

\bibitem[\protect\citeauthoryear{{Draine}}{{Draine}}{2011}]{Draine11}
{Draine} B.~T.,  2011, {Physics of the Interstellar and Intergalactic Medium}.
Princeton University Press

\bibitem[\protect\citeauthoryear{{Eikenberry} et~al.,}{{Eikenberry}
  et~al.}{2019a}]{Eikenberry19b}
{Eikenberry} S.,  et~al., 2019a, \baas, \href
  {https://ui.adsabs.harvard.edu/abs/2019BAAS...51g.124E} {51, 124}

\bibitem[\protect\citeauthoryear{{Eikenberry} et~al.,}{{Eikenberry}
  et~al.}{2019b}]{Eikenberry19c}
{Eikenberry} S.,  et~al., 2019b, \baas, \href
  {https://ui.adsabs.harvard.edu/abs/2019BAAS...51g.137E} {51, 137}

\bibitem[\protect\citeauthoryear{{Eikenberry} et~al.,}{{Eikenberry}
  et~al.}{2019c}]{Eikenberry19a}
{Eikenberry} S.,  et~al., 2019c, \baas, \href
  {https://ui.adsabs.harvard.edu/abs/2019BAAS...51c.283E} {51, 283}

\bibitem[\protect\citeauthoryear{{Flewelling} et~al.,}{{Flewelling}
  et~al.}{2020}]{Panstarrs}
{Flewelling} H.~A.,  et~al., 2020, \mn@doi [\apjs] {10.3847/1538-4365/abb82d},
  \href {https://ui.adsabs.harvard.edu/abs/2020ApJS..251....7F} {251, 7}

\bibitem[\protect\citeauthoryear{{Gaikwad}, {Srianand}, {Choudhury}  \&
  {Khaire}}{{Gaikwad} et~al.}{2017}]{Gaikwad17}
{Gaikwad} P.,  {Srianand} R.,  {Choudhury} T.~R.,   {Khaire} V.,  2017, \mn@doi
  [\mnras] {10.1093/mnras/stx248}, \href
  {https://ui.adsabs.harvard.edu/abs/2017MNRAS.467.3172G} {467, 3172}

\bibitem[\protect\citeauthoryear{{Hagen} et~al.,}{{Hagen}
  et~al.}{1992}]{Hagen92}
{Hagen} H.~J.,  et~al., 1992, \aap, \href
  {https://ui.adsabs.harvard.edu/abs/1992A&A...253L...5H} {253, L5}

\bibitem[\protect\citeauthoryear{{Hern{\'a}n-Caballero}, {Hatziminaoglou},
  {Alonso-Herrero}  \& {Mateos}}{{Hern{\'a}n-Caballero}
  et~al.}{2016}]{Hernan16}
{Hern{\'a}n-Caballero} A.,  {Hatziminaoglou} E.,  {Alonso-Herrero} A.,
  {Mateos} S.,  2016, \mn@doi [\mnras] {10.1093/mnras/stw2107}, \href
  {https://ui.adsabs.harvard.edu/abs/2016MNRAS.463.2064H} {463, 2064}

\bibitem[\protect\citeauthoryear{{Hu}, {Kim}, {Cowie}, {Songaila}  \&
  {Rauch}}{{Hu} et~al.}{1995}]{Hu95}
{Hu} E.~M.,  {Kim} T.-S.,  {Cowie} L.~L.,  {Songaila} A.,   {Rauch} M.,  1995,
  \mn@doi [\aj] {10.1086/117625}, \href
  {https://ui.adsabs.harvard.edu/abs/1995AJ....110.1526H} {110, 1526}

\bibitem[\protect\citeauthoryear{{Hui} \& {Rutledge}}{{Hui} \&
  {Rutledge}}{1999}]{Hui99}
{Hui} L.,  {Rutledge} R.~E.,  1999, \mn@doi [\apj] {10.1086/307202}, \href
  {https://ui.adsabs.harvard.edu/abs/1999ApJ...517..541H} {517, 541}

\bibitem[\protect\citeauthoryear{Ida, Ando  \& Toraya}{Ida
  et~al.}{2000}]{Ida00}
Ida T.,  Ando M.,   Toraya H.,  2000, \mn@doi [Journal of Applied
  Crystallography] {10.1107/S0021889800010219}, 33, 1311

\bibitem[\protect\citeauthoryear{{Jeram}, {Gonzalez}, {Eikenberry}, {Stern},
  {Mendes de Oliveira}, {Izuti Nakazono}  \& {Ackley}}{{Jeram}
  et~al.}{2020}]{Jeram20}
{Jeram} S.,  {Gonzalez} A.,  {Eikenberry} S.,  {Stern} D.,  {Mendes de
  Oliveira} C.~L.,  {Izuti Nakazono} L.~M.,   {Ackley} K.,  2020, \mn@doi
  [\apj] {10.3847/1538-4357/ab9c95}, \href
  {https://ui.adsabs.harvard.edu/abs/2020ApJ...899...76J} {899, 76}

\bibitem[\protect\citeauthoryear{{Jiao}, {Zhang}, {Zhang}, {Yu}, {Zhu}  \&
  {Li}}{{Jiao} et~al.}{2020}]{Jiao20}
{Jiao} K.,  {Zhang} J.-C.,  {Zhang} T.-J.,  {Yu} H.-R.,  {Zhu} M.,   {Li} D.,
  2020, \mn@doi [\jcap] {10.1088/1475-7516/2020/01/054}, \href
  {https://ui.adsabs.harvard.edu/abs/2020JCAP...01..054J} {2020, 054}

\bibitem[\protect\citeauthoryear{{Kim}, {Cristiani}  \& {D'Odorico}}{{Kim}
  et~al.}{2001}]{Kim01}
{Kim} T.~S.,  {Cristiani} S.,   {D'Odorico} S.,  2001, \mn@doi [\aap]
  {10.1051/0004-6361:20010650}, \href
  {https://ui.adsabs.harvard.edu/abs/2001A&A...373..757K} {373, 757}

\bibitem[\protect\citeauthoryear{{Kim}, {Linder}, {Edelstein}  \&
  {Erskine}}{{Kim} et~al.}{2015}]{Kim15}
{Kim} A.~G.,  {Linder} E.~V.,  {Edelstein} J.,   {Erskine} D.,  2015, \mn@doi
  [Astroparticle Physics] {10.1016/j.astropartphys.2014.09.004}, \href
  {https://ui.adsabs.harvard.edu/abs/2015APh....62..195K} {62, 195}

\bibitem[\protect\citeauthoryear{{Kloeckner} et~al.,}{{Kloeckner}
  et~al.}{2015}]{Kloeckner15}
{Kloeckner} H.~R.,  et~al., 2015, Advancing Astrophysics with the Square
  Kilometre Array (AASKA14), \href
  {https://ui.adsabs.harvard.edu/abs/2015aska.confE..27K} {p.~27}

\bibitem[\protect\citeauthoryear{{Kuhr}, {Liebert}, {Strittmatter}, {Schmidt}
  \& {Mackay}}{{Kuhr} et~al.}{1983}]{Kuhr83}
{Kuhr} H.,  {Liebert} J.~W.,  {Strittmatter} P.~A.,  {Schmidt} G.~D.,
  {Mackay} C.,  1983, \mn@doi [\apjl] {10.1086/184166}, \href
  {https://ui.adsabs.harvard.edu/abs/1983ApJ...275L..33K} {275, L33}

\bibitem[\protect\citeauthoryear{{Liddle}}{{Liddle}}{2007}]{Liddle07}
{Liddle} A.~R.,  2007, \mn@doi [\mnras] {10.1111/j.1745-3933.2007.00306.x},
  \href {https://ui.adsabs.harvard.edu/abs/2007MNRAS.377L..74L} {377, L74}

\bibitem[\protect\citeauthoryear{{Liske} et~al.,}{{Liske}
  et~al.}{2008}]{Liske08}
{Liske} J.,  et~al., 2008, \mn@doi [\mnras] {10.1111/j.1365-2966.2008.13090.x},
  \href {https://ui.adsabs.harvard.edu/abs/2008MNRAS.386.1192L} {386, 1192}

\bibitem[\protect\citeauthoryear{{Loeb}}{{Loeb}}{1998}]{Loeb98}
{Loeb} A.,  1998, \mn@doi [\apjl] {10.1086/311375}, \href
  {https://ui.adsabs.harvard.edu/abs/1998ApJ...499L.111L} {499, L111}

\bibitem[\protect\citeauthoryear{{Logsdon} et~al.,}{{Logsdon}
  et~al.}{2018}]{Logsdon18}
{Logsdon} S.~E.,  et~al., 2018, in {Evans} C.~J.,  {Simard} L.,   {Takami} H.,
  eds,  Society of Photo-Optical Instrumentation Engineers (SPIE) Conference
  Series Vol. 10702, Ground-based and Airborne Instrumentation for Astronomy
  VII. p. 1070267, \mn@doi{10.1117/12.2312209}

\bibitem[\protect\citeauthoryear{{Mayor} et~al.,}{{Mayor}
  et~al.}{2003}]{Mayor03}
{Mayor} M.,  et~al., 2003, The Messenger, \href
  {https://ui.adsabs.harvard.edu/abs/2003Msngr.114...20M} {114, 20}

\bibitem[\protect\citeauthoryear{{McVittie}}{{McVittie}}{1962}]{McVittie62}
{McVittie} G.~C.,  1962, \apj, \href
  {https://ui.adsabs.harvard.edu/abs/1962ApJ...136..334M} {136, 334}

\bibitem[\protect\citeauthoryear{{Morris} et~al.,}{{Morris}
  et~al.}{2018}]{Morris18}
{Morris} B.~M.,  et~al., 2018, \mn@doi [\aj] {10.3847/1538-3881/aaa47e}, \href
  {https://ui.adsabs.harvard.edu/abs/2018AJ....155..128M} {155, 128}

\bibitem[\protect\citeauthoryear{{Newville}, {Stensitzki}, {Allen}  \&
  {Ingargiola}}{{Newville} et~al.}{2014}]{lmfit}
{Newville} M.,  {Stensitzki} T.,  {Allen} D.~B.,   {Ingargiola} A.,  2014,
  {LMFIT: Non-Linear Least-Square Minimization and Curve-Fitting for Python},
  \mn@doi{10.5281/zenodo.11813}

\bibitem[\protect\citeauthoryear{{Pepe} et~al.,}{{Pepe} et~al.}{2002}]{Pepe02}
{Pepe} F.,  et~al., 2002, The Messenger, \href
  {https://ui.adsabs.harvard.edu/abs/2002Msngr.110....9P} {110, 9}

\bibitem[\protect\citeauthoryear{{Perlmutter} et~al.,}{{Perlmutter}
  et~al.}{1999}]{Perlmutter99}
{Perlmutter} S.,  et~al., 1999, \mn@doi [\apj] {10.1086/307221}, \href
  {https://ui.adsabs.harvard.edu/abs/1999ApJ...517..565P} {517, 565}

\bibitem[\protect\citeauthoryear{{Planck Collaboration} et~al.,}{{Planck
  Collaboration} et~al.}{2016}]{Planck16}
{Planck Collaboration} et~al., 2016, \mn@doi [\aap]
  {10.1051/0004-6361/201525830}, \href
  {https://ui.adsabs.harvard.edu/abs/2016A&A...594A..13P} {594, A13}

\bibitem[\protect\citeauthoryear{{Riess} et~al.,}{{Riess}
  et~al.}{1998}]{Riess98}
{Riess} A.~G.,  et~al., 1998, \mn@doi [\aj] {10.1086/300499}, \href
  {https://ui.adsabs.harvard.edu/abs/1998AJ....116.1009R} {116, 1009}

\bibitem[\protect\citeauthoryear{{Sandage}}{{Sandage}}{1962}]{Sandage62}
{Sandage} A.,  1962, \mn@doi [\apj] {10.1086/147385}, \href
  {https://ui.adsabs.harvard.edu/abs/1962ApJ...136..319S} {136, 319}

\bibitem[\protect\citeauthoryear{{Schindler}, {Fan}, {McGreer}, {Yang}, {Wu},
  {Jiang}  \& {Green}}{{Schindler} et~al.}{2017}]{Schindler17}
{Schindler} J.-T.,  {Fan} X.,  {McGreer} I.~D.,  {Yang} Q.,  {Wu} J.,  {Jiang}
  L.,   {Green} R.,  2017, \mn@doi [\apj] {10.3847/1538-4357/aa9929}, \href
  {https://ui.adsabs.harvard.edu/abs/2017ApJ...851...13S} {851, 13}

\bibitem[\protect\citeauthoryear{{Schindler} et~al.,}{{Schindler}
  et~al.}{2018}]{Schindler18}
{Schindler} J.-T.,  et~al., 2018, \mn@doi [\apj] {10.3847/1538-4357/aad2dd},
  \href {https://ui.adsabs.harvard.edu/abs/2018ApJ...863..144S} {863, 144}

\bibitem[\protect\citeauthoryear{{Schindler} et~al.,}{{Schindler}
  et~al.}{2019a}]{Schindler19b}
{Schindler} J.-T.,  et~al., 2019a, \mn@doi [\apjs] {10.3847/1538-4365/ab20d0},
  \href {https://ui.adsabs.harvard.edu/abs/2019ApJS..243....5S} {243, 5}

\bibitem[\protect\citeauthoryear{{Schindler} et~al.,}{{Schindler}
  et~al.}{2019b}]{Schindler19a}
{Schindler} J.-T.,  et~al., 2019b, \mn@doi [\apj] {10.3847/1538-4357/aaf86c},
  \href {https://ui.adsabs.harvard.edu/abs/2019ApJ...871..258S} {871, 258}

\bibitem[\protect\citeauthoryear{{Schmidt} et~al.,}{{Schmidt}
  et~al.}{1998}]{Schmidt98}
{Schmidt} B.~P.,  et~al., 1998, \mn@doi [\apj] {10.1086/306308}, \href
  {https://ui.adsabs.harvard.edu/abs/1998ApJ...507...46S} {507, 46}

\bibitem[\protect\citeauthoryear{{Schwab}, {St{\"u}rmer}, {Gurevich},
  {F{\"u}hrer}, {Lamoreaux}, {Walther}  \& {Quirrenbach}}{{Schwab}
  et~al.}{2015}]{Schwab15}
{Schwab} C.,  {St{\"u}rmer} J.,  {Gurevich} Y.~V.,  {F{\"u}hrer} T.,
  {Lamoreaux} S.~K.,  {Walther} T.,   {Quirrenbach} A.,  2015, \mn@doi [\pasp]
  {10.1086/682879}, \href
  {https://ui.adsabs.harvard.edu/abs/2015PASP..127..880S} {127, 880}

\bibitem[\protect\citeauthoryear{{Schwab} et~al.,}{{Schwab}
  et~al.}{2016}]{Schwab16}
{Schwab} C.,  et~al., 2016, in {Evans} C.~J.,  {Simard} L.,   {Takami} H.,
  eds,  Society of Photo-Optical Instrumentation Engineers (SPIE) Conference
  Series Vol. 9908, Ground-based and Airborne Instrumentation for Astronomy VI.
  p. 99087H, \mn@doi{10.1117/12.2234411}

\bibitem[\protect\citeauthoryear{{Vogt} et~al.,}{{Vogt} et~al.}{1994}]{Vogt94}
{Vogt} S.~S.,  et~al., 1994, in {Crawford} D.~L.,  {Craine} E.~R.,  eds,
  Society of Photo-Optical Instrumentation Engineers (SPIE) Conference Series
  Vol. 2198, Instrumentation in Astronomy VIII. p.~362,
  \mn@doi{10.1117/12.176725}

\bibitem[\protect\citeauthoryear{{Wilken} et~al.,}{{Wilken}
  et~al.}{2012}]{Wilken12}
{Wilken} T.,  et~al., 2012, \mn@doi [\nat] {10.1038/nature11092}, \href
  {https://ui.adsabs.harvard.edu/abs/2012Natur.485..611W} {485, 611}

\bibitem[\protect\citeauthoryear{{Wolf} et~al.,}{{Wolf} et~al.}{2020}]{Wolf20}
{Wolf} C.,  et~al., 2020, \mn@doi [\mnras] {10.1093/mnras/stz2955}, \href
  {https://ui.adsabs.harvard.edu/abs/2020MNRAS.491.1970W} {491, 1970}

\bibitem[\protect\citeauthoryear{{Zhang}, {Meiksin}, {Anninos}  \&
  {Norman}}{{Zhang} et~al.}{1998}]{Zhang98}
{Zhang} Y.,  {Meiksin} A.,  {Anninos} P.,   {Norman} M.~L.,  1998, \mn@doi
  [\apj] {10.1086/305260}, \href
  {https://ui.adsabs.harvard.edu/abs/1998ApJ...495...63Z} {495, 63}

\makeatother
\end{thebibliography}

\appendix

\section{Voigt profile}\label{sec:voigteqs}
The Voigt line profile is the intrinsic Lorentz line profile convolved with the Gaussian velocity distribution of the absorbers \citep[][Eq. 6.37]{Draine11}:
\begin{equation}\label{eq:voigt}
    \phi_\nu = \frac{4\gamma}{b\sqrt{\pi}}
    \int_{-\infty}^{+\infty}
    \frac{\exp\left(-\frac{v^2}{b^2}\right)}{16\pi^2\left(\nu-\nu_0+\frac{\nu_0}{c}v\right)^2+\gamma^2}\dd v,
\end{equation}
where $b \equiv \sqrt{2}\sigma_v$ is the Doppler broadening parameter, $\gamma$ is the Lorentz line width parameter. For Ly$\alpha$, $\gamma = 7618\lambda^{-1}~\mathrm{cm~s}^{-1}$ \citep[][p. 58]{Draine11}.
Equation \ref{eq:voigt} can be rewritten as 
$\phi_\nu
    = \frac{c}{\sqrt{\pi}\nu_0b^2} K(x, y)$,
where
\begin{equation}
K(x, y) = \frac{y}{\pi} \int_{-\infty}^{+\infty}
    \frac{e^{-t^2}}{(t - x)^2 + y^2} \dd t,
\end{equation}
$x = \frac{c(\nu_0-\nu)}{\nu_0b}$, $y = \frac{\gamma c}{4\pi\nu_0b}$ and $t = v/b$.
By approximation, $K(x, y) \approx \mathbb{R}[w(z)]$ \citep[e.g.][]{Ida00}, where $w(z) = e^{-z^2}\mathrm{erfc}(-iz)$ is the Faddeeva function, $z = x + iy$.
Given the normalized Voigt profile $\phi_\nu$, the optical depth
\begin{gather}
\tau_\nu = \frac{\pi e^2}{m_e c} fN_{\mathrm{HI}}\phi_\nu,
\end{gather}
where $f = 0.4164$ is the oscillator strength for Ly$\alpha$, $N_{\mathrm{HI}}$ is the HI column density.
Finally, the normalized spectrum can be generated by combining each parameterized line $i$ by
$S_\nu = \exp\left(-\sum_i \tau_{\nu,i}\right)$.

\section{Southern sample}\label{sec:southern}
In addition to the northern sample of optimal targets for the redshift drift experiment, here we present the corresponding southern sample. The sample is selected by restricting the declinations of the targets to be less than $+20^\circ$. Table~\ref{tab:southern} lists the best 10 southern quasars selected by $|\dot{v}|/\sigma_{\dot{v}}$.

\begin{table*}
\caption{Southern sample selected by $|\dot{v}|/\sigma_{\dot{v}}$.}
\label{tab:southern}
\begin{tabular}{clrrccc}
\hline
Rank & Name & RA (deg)&     DEC (deg) & $M_\mathrm{1450}$ & $r$ (mag) & $z$\\
\hline
1  &  PS1 J212540.96-171951.4 &  321.420691 & -17.330947 & -29.35 &  16.50 &  3.900 \\
2  & SDSS J034151.16+172049.7 &   55.463213 &  17.347164 & -29.59 &  16.26 &  3.690 \\
3  & SMSS J054803.19-484813.1 &   87.013348 & -48.803654 &    --- &  17.22 &  4.147 \\
4  &  PS1 J052136.92-133938.8 &   80.403846 & -13.660775 & -28.36 &  17.68 &  4.270 \\
5  &                 Q0000-26 &    0.845642 & -26.055084 & -28.73 &  17.44 &  4.111 \\
6  & SDSS J095937.11+131215.5 &  149.904628 &  13.204312 & -28.51 &  17.39 &  4.064 \\
7  &  PS1 J111054.69-301129.9 &  167.727862 & -30.191653 & -28.87 &  18.64 &  4.830 \\
8  &  PS1 J222152.88-182602.9 &  335.470344 & -18.434150 & -28.30 &  18.19 &  4.520 \\
9  &             BR 2237-0607 &  339.973672 &  -5.872222 & -28.31 &  18.25 &  4.550 \\
10 &  PS1 J010318.08-130510.2 &   15.825312 & -13.086163 & -28.61 &  17.46 &  4.065 \\
\hline
\end{tabular}
\end{table*}

\begin{figure*}
    \includegraphics[width=\columnwidth]{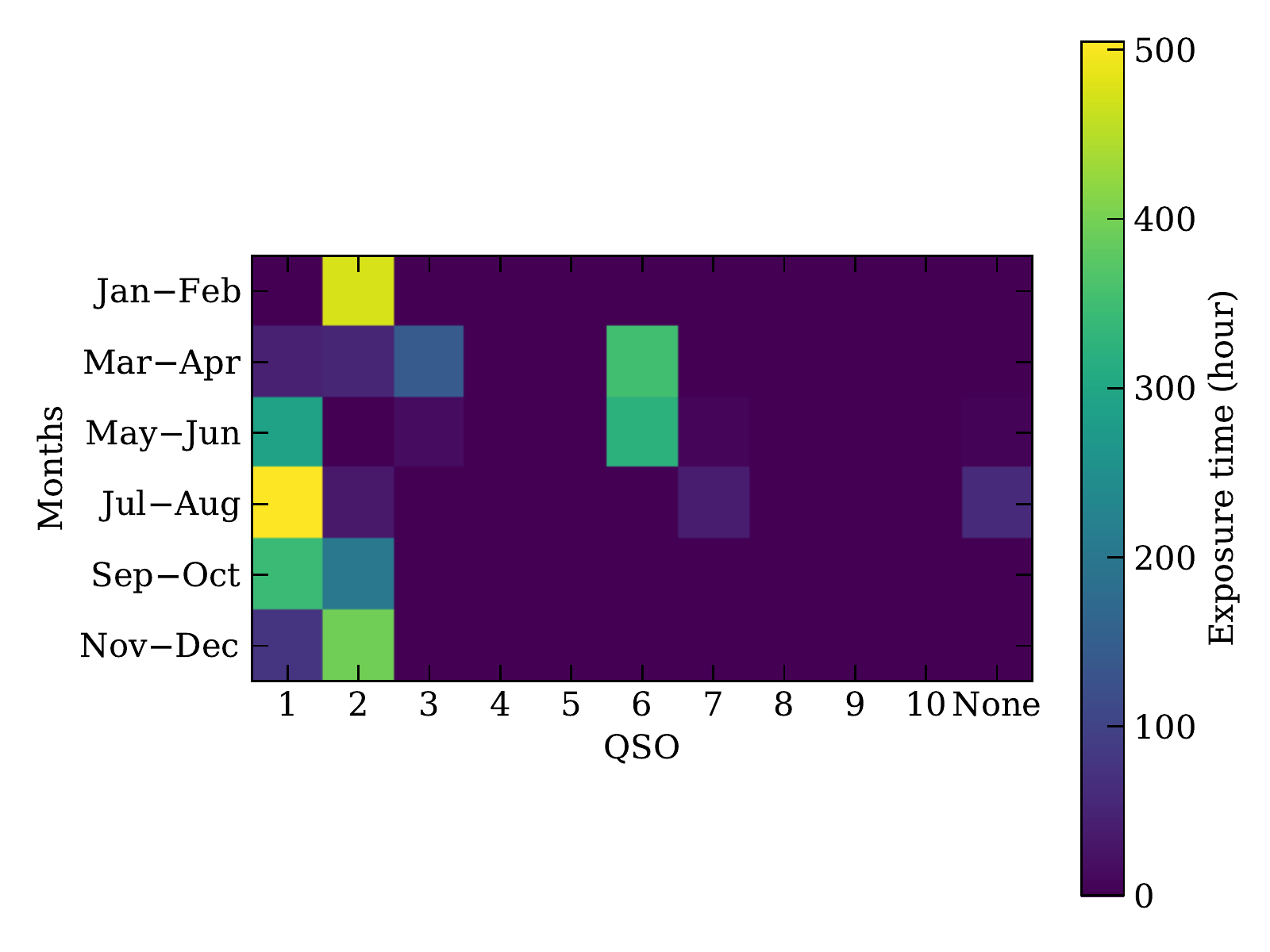}
    \includegraphics[width=\columnwidth]{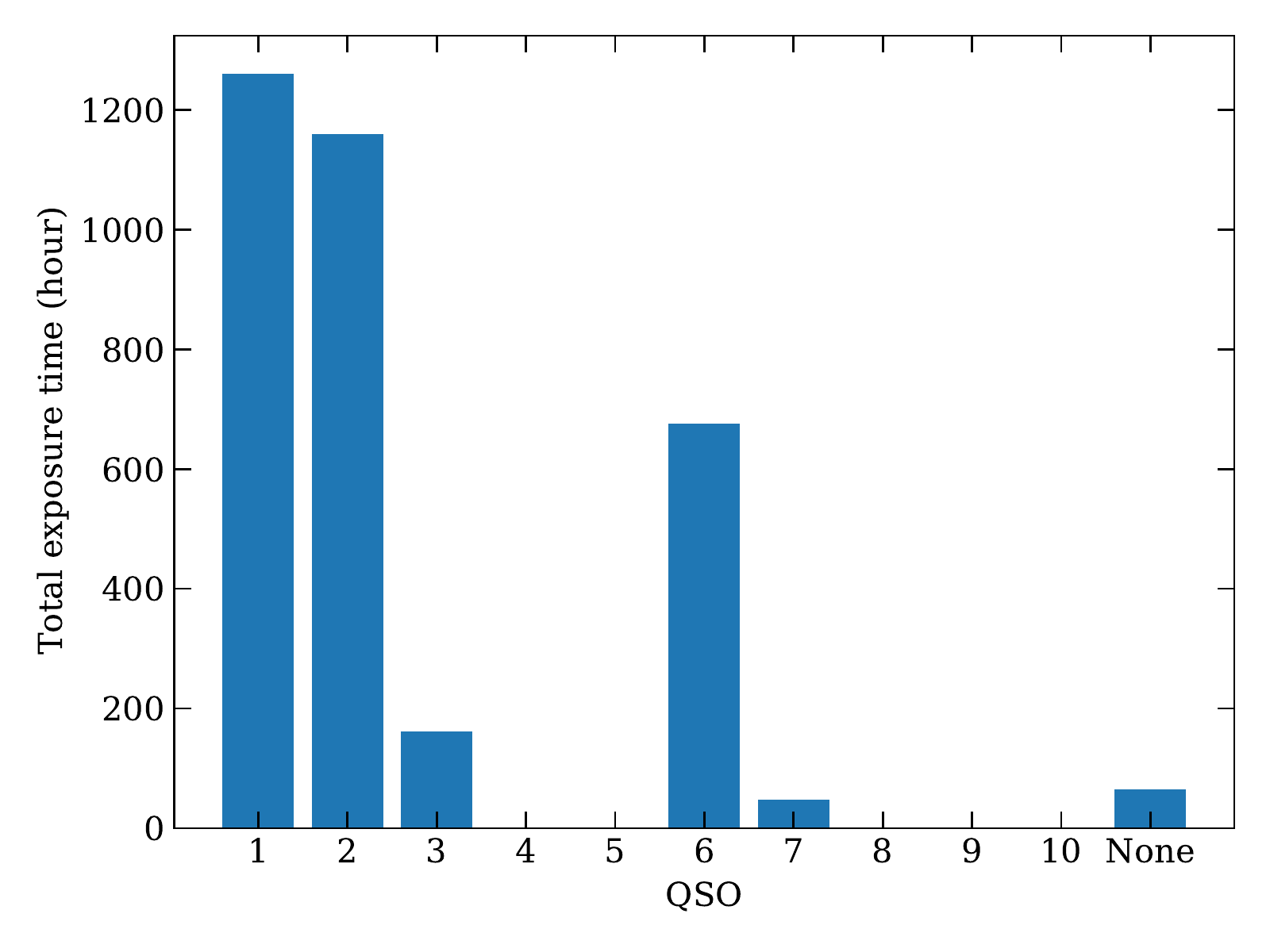}
    \caption{Same as Figure~\ref{fig:time} but for the best 10 quasars in the southern sample for a planned experiment at the Paranal Observatory.
     The QSO numbers correspond to the rank of the southern sample with respect to the expected $|\dot{v}|/\sigma_{\dot{v}}$ (as listed in the first column of Table~\ref{tab:southern}), i.e. the observing priority. ``None'' corresponds to time when none of these 10 QSOs is observable. This figure indicates that for the southern sample, after considering 10 optimal quasars, there is still time left when no target is visible.
    }\label{fig:southern}
\end{figure*}

We also planned the observations of each night over the year 2021, as described in Section \ref{sec:time}, assuming the telescope is at the Paranal Observatory ($24^\circ37'39''$ S, $70^\circ24'18''$ W). Figure~\ref{fig:southern} shows the exposure time distribution among the best 10 quasars in the southern sample in one year. 
Unlike the northern sample, which utilizes all night time with only 4 targets, there is still 64.3 hours of time left after considering all 10 targets. 
There are 2581.4 hours (76.6\%) of time allocated to the best 3 quasars, which is much lower than the corresponding fraction for the northern sample (97.7\%).
Therefore, for the currently known QSOs, the northern hemisphere yields a higher S/N for a given observing time. This conclusion should be re-evaluated after Gaia DR3 to incorporate any new high-redshift quasars discovered by that mission.

\bsp
\label{lastpage}
\end{CJK*}
\end{document}